\documentclass[10pt]{article}
\usepackage{fullpage}
\usepackage{spec}
\usepackage{epsfig}
\usepackage{epsf}
\usepackage{amsmath}
\usepackage[footnotesize]{caption}
\usepackage{setspace}

\newcommand{\bm}[1]{\mbox{\boldmath $#1$}}
\newcommand{\mb}[1]{\mathbf{#1}}
\newcommand{\color}[1]{\bm}

\begin{document}
\title{Microcolony and Biofilm Formation\\ as a Survival Strategy for Bacteria}
\author{
Leah R. Johnson$^{1*}$\\
{\small $^1$ Department of Physics}\\
{\small University of California, Santa Cruz, 1156 High St., Santa Cruz, CA 95064}\\
{\small Current address:  Statistical Laboratory / CMS}\\
{\small Wilberforce Road, Cambridge
CB3 0WB, United Kingdom.}  \\{\small E-mail: {\tt leah@statslab.cam.ac.uk}}
\date{}
}
\maketitle


\begin{abstract}
  Bacterial communities such as biofilms are widely recognised as
  being important for survival and persistence of bacteria in harsh
  environments. Mechanistic models of biofilm growth indicate that the
  way in which the surface is seeded can effect the morphology of
  simulated biofilms. Experimental studies indicate that genes which
  are important for chemotaxis also influence biofilm formation,
  perhaps by influencing aggregation on a surface. Understanding
  aggregation and microcolony formation could therefore help clarify
  factors influencing biofilm formation as well as understanding how
  groups may influence the fitness of bacteria. In this paper I
  develop an Individual Based Model to examine how different behaviours
  involved in microcolony formation on a surface determine patterns
  of group sizes, and link patterns to bacterial fitness. I also
  provide a method for comparing data with model hypotheses to
  identify bacterial behaviours in experimental systems.
\end{abstract}

\section{Introduction}
Bacteria live in communities for many of the same reasons that other
organisms live in communities; for example protection from predators
or other external dangers and access to resources, and genetic
diversity \cite{jefferson:2004,roszak:1987}. Individuals in many
bacterial communities, such as in biofilms, experience increased
resistance to antibiotics, thermal stress, or predation
\cite{hahn:2001,matz:2005}.  These communities also allow bacteria to
stay in favourable environments without being swept away. Bacterial
fitness depends strongly on how quickly the cells can double. However,
because doubling rates of individuals in a community are generally
lower than doubling rates of individuals outside of communities,
living in a community often represents a trade-off between
reproduction and survival.

Community types and morphologies are determined by physical,
biochemical, and genetic factors. Physical and biochemical factors --
such as fluid flow, polymer tensile strength, and cell surface
properties -- influence the availability of nutrients within a
community, the ability of a biofilm to hold together under shear
stress, and determine whether or not cells can stick to a particular
surface. Genetic factors constrain the behavioural options available to
the bacteria, and determine responses to chemical or environmental
signals. Current research indicates that gene expression of bacteria
living in biofilms or other communities differs significantly from
that of free-living cells, and may be mediated by a process called
quorum sensing \cite{davies:1998,heydorn:2002,parsek:2005}.

Genes that regulate the production of extra-cellular polymeric
substances (EPS) are widely recognised as influencing biofilm
formation
\cite{branda:2005,davey:2000,mayer:1999,yildiz:1999,watnick:1999}.
Current research indicates that genes involved in flagellar motility
and various chemotaxis and quorum sensing systems also appear to also
influence the ability of many types of bacteria, for instance
toxigenic {\it Vibrio cholerae}, to form biofilms, as well as
influencing the morphology of the biofilms
\cite{davies:1998,heydorn:2002,parsek:2005,suntharalingam:2005}. Since
many bacteria in biofilms appear to down-regulate motility
genes, it has been hypothesised that aggregation of bacteria on a
surface is an important step in the development of biofilms
\cite{watnick:1999}. However, data on the role of aggregation in
initial stages of microcolony formation are inconclusive
\cite{klausen:2003}. This is partly due to the fact that bacterial
aggregation is difficult to observe directly. Observations of
densities are influenced by both aggregation and reproduction. It can
also be very difficult to monitor individual bacteria on a surface,
since many microscopic techniques that are able to clearly discern
individual cells, such as confocal or scanning electron microscopy,
often damage or kill cells.

Since aggregation can be hard to observe directly, mathematical models
that can predict observable macroscopic patterns indicative of
aggregation are desirable.  Models also allow exploration of how
bacterial behaviours change macroscopic patterns, and help in linking
these behaviours to the function of various genes.

In this paper I introduce an Individual Based Model (IBM) for
microcolony formation on a surface. I seek to addresses three major
questions. First, how do interactions between cells and cell behaviours
influence microcolony characteristics such as size? Second, can we
differentiate between different kinds of behaviours from ``snapshots''
of growing microcolonies? Finally, how might these behaviours and the
resulting patterns be linked to bacterial fitness?

\section{Approaches for Modelling Bacterial Communities} \label{literature:colony}

Models of communities of micro-organisms, as well as larger organisms,
generally take one of three approaches. The first approach to is
generally referred to as Eulerian methods. These are continuous
models, focused on developing field equations that describe the flux
of individuals in space. This type of approach ignores individual
identity, instead focusing on densities of individuals in an area or
volume. Examples of this approach include Keller and Segel's models of
slime mold aggregation \cite{keller:1970} and bacterial chemotaxis
\cite{keller:1971a,keller:1971b}, models of midge swarming
\cite{tyutyunov:2004}, and a general approach to group formation in
animals proposed by Gueron and Levin \cite{gueron:1995}.  The Eulerian
approach is most useful for describing very large numbers of
individuals with high density. It also has an advantage that there are
many good analytical tools available. However, if the density of
individuals is low, or one wishes to explore how individual strategies
effect group dynamics, this type of approach is less useful
\cite{gueron:1996}.

A second approach uses Cellular Automata (CA) \cite{wolfram}. CA are a
class of discrete dynamical models where the space occupied is divided
into grids or boxes (often called cells, but for our purpose here they
will be referred to as {\it elements}) on a lattice. Each element is
characterised by some description of the element's state at some time
$t$, which is often discrete. The state of each element can be updated
by a set of rules that may depend on the current state of the element
as well as the states of neighbour elements. This method is especially
popular for modelling biofilm formation. For models of biofilms, the
state of each element contains information about whether the cell is
occupied by biomass of particular density and type. Such models are
sometimes referred to as biomass-based models (BbM) \cite{kreft:2001}.

Some of the most detailed biofilm models, developed by Picioreanu {\it
  {\it et al.}}
\cite{picioreanu:1997,picioreanu2,picioreanu,vanLoosdrecht:2002},
involve a CA model for the spreading of biomass, including living and
dead cells and extra-cellular polymeric substances (EPS).
Additionally, these methods include substrate transport via diffusion
and convection, and can include modelling of detachment due to
mechanical stress from fluid flow over the biofilm.

Although both Eulerian and CA models can provide information about
macroscopic properties of systems, they ignore both variation of
traits between individuals and interactions between individuals. Since
both of these can greatly effect the dynamics of the larger system
that the individuals compose, a third approach, Individual Based
Models (IBMs) \cite{grimm} can be used to explore these phenomena
explicitly. IBMs comprise models where the behaviour of each individual
is modelled separately, and each individual follows a set of rules that
determine their
behaviour.

IBM models are variants of $N$-body Newtonian dynamics problems, and
are often called Lagrangian models. The rules that control individuals
are a combination of forces and decision rules. Examples of forces
include those that are physical or environmental (such as chemical
gradients, gravity, or drag), and other such as attraction or
repulsion between individuals \cite{okubo:1986,gueron:1996}.
Regardless of the types of decisions or forces chosen, the goal is to
learn about how interactions and behaviours on small scales influence
large-scale patterns. Additionally, once behaviours that result in
``realistic'' patterns have been identified, these can be compared to
behaviours in populations, and used to evaluate which rules could be
selected for in different situations.

BacSim \cite{kreft:1998,kreft:2001}, a model of {\it E. coli} colony
growth, is an example of an IBM model of biofilm growth. It features
an IBM for the growth and behaviour of individual bacteria (including
uptake of substrate, death, and reproduction) together with a
simulation of the diffusion and reaction of substrate and other
products. BacSim simulates spreading of the biomass by requiring that
a minimum distance is maintained between cells.  One of the primary
results of the study by Kreft {\it {\it et al.}}  \cite{kreft:2001} is
that the initial seeding of a surface has a major impact on the
development and morphology of simulated biofilms, especially those
composed of multiple types of bacteria. This is likely because of the
heterogeneity of substrate concentration in the biofilm. They
concluded that ``this stresses the primary importance of spreading and
chance in the emergence of the complexity of the biofilm community.''

Biofilm formation is generally thought to proceed as follows:
	\begin{enumerate}
	\item individuals colonise surface
	\item individuals form microcolonies 
	\item microcolonies form biofilms
	\end{enumerate}
   
Most current models of biofilms, regardless of approach, focus on
growth of biofilms from a randomly seeded surface, using physical and
chemical factors, i.e., they focus on the third step listed above.
These models ignore the initial surface colonisation events and
microcolony formation. They also largely ignore biochemical
interactions, such as quorum sensing. 

My goal is to develop an IBM that can be used with experimental data
to infer the bacterial behaviours that generate the patterns observed
during the initial stages of biofilm formation. I am most interested
in how interactions among individuals affect microcolony sizes during
initial stages of biofilm formation, and the implications the
behaviours have for bacterial fitness`. I assume that the environment is
homogeneous and external forces (such as drag) are minimal. I also
assume that the density of organisms is sufficiently low so that
growth and reproduction are not affected by diffusion limitation, and
competition for nutrients is minimal.

\section{An IBM for Bacterial Community Formation} \label{math_model}

\subsection{Cell State}\label{cellstate}
The IBM begins with a description of the variables necessary to
describe an individual cell's state. At time $t$ a cell is
characterised by a set of state variables. These variables can include
the cell's position, velocity, genotype, age, size, {\it etc.}  I will
denote the vector that describes the overall state of the $i^{th}$
cell as $\mb{S}_i(t)$. For the models presented here, the state of the
cell is described by a combination of the variables listed in Table
\ref{tb:statevars1}.

\begin{table}
\begin{center}
\begin{tabular}{| c | c | }
\hline
Symbol & Description  \\
\hline
$\mb{X}_i$ & cell position (centre of cell) \\
$\mb{F}_i$ & force on cell \\
$\mb{V}_i$ & velocity of cell \\
$\tau_i$ & time since last division\\
$T_i$ & doubling time of cell \\
$a_i$ & cellular age, in number of divisions  \\
$\gamma_i$ & cell is moving $\gamma_i = 1$, or stopped $\gamma_i = 0$ \\
$\alpha_i$ & cell is alive $\alpha_i = 1$, or dead $\alpha_i = 0$ \\
$\phi_i$ & cell overlaps other cells $\phi_i = 0$, otherwise $\phi_i = 1$ \\
$r_i$ & cell radius \\
$C_i$ & colony membership of the cell \\
\hline
\end{tabular} 
\caption[State variables included in initial models]{State Variables for the $i^{th}$ cell, included in models explored in this chapter. \label{tb:statevars1}}
\end{center}
\end{table}

\begin{figure}[ht!]
\begin{center}
\includegraphics[scale=0.5]{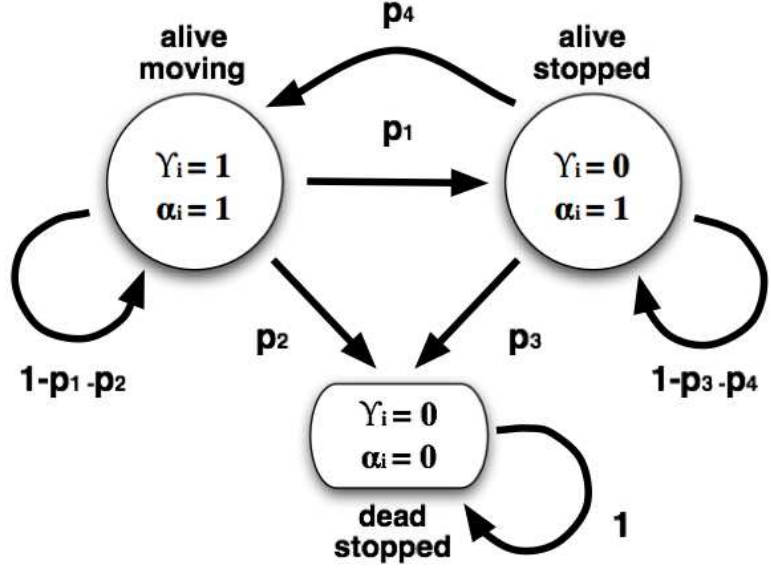} \\
\caption[Schematic of transitions between stage states for cell
$i$]{\label{plot:colonystate1} Schematic of transitions between
  stage-states for cell $i$, without interactions with other cells.
  Transition probabilities are denoted by $p_1$ through $p_4$}
\end{center}
\end{figure}

I denote the {\it stage-state} of the $i^{th}$cell at time $t$ by
$s_{i,t}=\{a, \gamma, \alpha, \phi\}_{i,t}$. The value of the
stage-state is the component of the full state of the cell,
$\mb{S}_i(t)$, that describes whether the cell is moving and whether
it is alive. The cell can then transition between various
stage-states, and the full states, at any time step. Figure
\ref{plot:colonystate1} shows a graphical representation of the
conditional transitions between the various stages and their
probabilities, assuming the transitions are age independent. In a
small increment of time from $t$ to $t+\Delta t$, a moving cell stops
with probability $p_1$. A moving cell can also die (and thus
simultaneously stop) with probability $p_2$. A stopped cell can die
with probability $p_3$. Finally, a stopped but alive cell can move
again with probability $p_4$.  In other words:
\begin{align*}
Pr\{ \gamma_{i,t+\Delta t} =0, \alpha_{i,t+\Delta t}=1\; | \; \gamma_{i,t}=1, \alpha_{i,t}=1\}&= p_1 \\
Pr\{ \gamma_{i,t+\Delta t}=0, \alpha_{i,t+\Delta t}=0 \; | \;\gamma_{i,t}=1, \alpha_{i,t}=1\}&= p_2\\
Pr\{ \gamma_{i,t+\Delta t}=0, \alpha_{i,t+\Delta t}=0 \; | \; \gamma_{i,t}=0, \alpha_{i,t}=1\}&= p_3\\
Pr\{ \gamma_{i,t+\Delta t}=1, \alpha_{i,t+\Delta t}=1 \; | \;\gamma_{i,t}=0, \alpha_{i,t}=1\}&= p_4 \\
\end{align*}
We can also describe this system using a transition matrix, in order
to connect the state of the $i^{th}$ cell at time $t+\Delta t$ to
the state at time $t$:
\begin{equation}
\left[ \begin{array}{c}
\gamma_{i,t+\Delta t}=1, \alpha_{i,t+\Delta t}=1 \\
\gamma_{i,t+\Delta t}=0, \alpha_{i,t+\Delta t}=1  \\
\gamma_{i,t+\Delta t}=0, \alpha_{i,t+\Delta t}=0 \\
\end{array} \right] = 
\left[ \begin{array}{ccc}
1-p_1-p_2 	& p_4		& 0	\\
p_1			& 1-p_3-p_4	& 0	\\
p_2			& p_3		& 1	\\
\end{array} \right]
\left[ \begin{array}{c}
\gamma_{i,t}=1, \alpha_{i,t}=1 \\
\gamma_{i,t}=0, \alpha_{i,t}=1  \\
\gamma_{i,t}=0, \alpha_{i,t}=0 \\
\end{array} \right] ,
\end{equation}
with the first matrix on the r.h.s. corresponding to the transition
matrix.  In the following analysis, I assume that the probability of
stopping is $p_1=P_1$, all cells die with equal probability,
($p_2=p_3=P_2$), and no stopped cells can start moving again ($p_4 =
0$).

\subsection{Cell movement} \label{cellmove}

In the initial stages of surface colonisation, bacteria first move
through a medium until they encounter and attach to a surface. This
movement can be described as a series of runs and tumbles due to
flagellar action \cite{brock}. A bacterium ``runs'' in a straight line for short
distances, and then ``tumbles'' to change the direction of travel. If
there are no external stimuli, this motion is random; directions of
runs will be uncorrelated. A bacterium may move on a surface, by
flagellar or twitching motility, and may exhibit a run and tumble
behaviour in this case as well. 

A bacterium's direction and speed of movement through a medium or on a
surface is influenced by various stimuli. The collective response to
stimuli is known as taxis. For instance, chemotaxis is movement in
response to chemical gradients, and phototaxis is movement in response
to light gradients. Since bacteria are very small, they explore
gradients with movement, keeping track of the temporal changes in
signal strength instead of measuring differences in concentrations of
a signal across the cell body. There is evidence that a bacterium
experiencing temporal increases in an attractant decreases the
frequency of tumbles, resulting in longer runs when moving up the
chemical gradient. When the attractant level decreases, tumbling
frequency increases, giving the bacterium more opportunities to find
the desired direction.

In this model, $N$ cells, denoted by $i=1, \dots, N$, are confined to
a two dimensional surface of area $A$. Each cell is modelled as a
circle of radius $r$ and all cells are of equal size, {\it i.e.},
$r_i=r_j=r$ $\forall$ $i, j$. As a baseline description, I assume that
in the absence of any interactions between the cells, each cell's
movement can be described as an Ornstien-Uhlenbeck process:
\begin{eqnarray*}
d{\bf X}_i &=& {\bf V}_i dt\\
d{\bf V}_i &=& -\eta {\bf V}_i dt + q d{\bf W}.
\end{eqnarray*}
Here $\eta$ is a drag or dissipation coefficient. However, in the rest
of this paper I will assume that $\eta = 0$, so that the velocities of
bacteria to not dissipate. This is reasonable as bacteria can propel
themselves, and overcome viscous forces. In this model, a cell moves
with a random velocity at each time step, with $dW \sim N(0,1)$, and
$q$ determining the change in the velocity at each time step. In this
case, cells will remain randomly distributed in the space without clumping
developing over time.

Next, I add simple interactions between cells and the surface. First,
if two cells bump into each other ({\it i.e.} if $|{\bf X}_i - {\bf
X}_j| \leq 2r$) they stick to each other and to the surface, and stop
moving. This is indicated using the state variable $\phi_i$ which is
defined as
\begin{equation} \label{eq:phi}
\phi_i = \prod_{\stackrel{j=0}{j \ne i}}^N H(|{\bf X}_i - {\bf X}_j|-2r),
\end{equation}
where $H(x)$ is the Heaviside step function\footnote{The Heaviside
function is a discontinuous step function, defined as
\begin{equation*}
H(x) = \begin{cases}
	0 & x < 0 \\
	1 & x \ge 0.
	\end{cases}
\end{equation*}
}, so that $\phi_i=1$ if the $i$th cell is not overlapping any other
cells, and $\phi_i=0$ otherwise.

I also allow individual cells to stop and stick to the surface with
non-zero probability $P_1$, independently of interactions with other
cells. The state variable $\gamma_{i,t}$ indicates if the cell is
moving: $\gamma_{i,t}=1$ when the cell is moving, and $\gamma_{i,t}=0$
when the cell has stopped. Also, if $\phi_i=0 \Rightarrow
\gamma_{i,t}=0$ Then $\gamma_{i,t}=1 \rightarrow \gamma_{i,t+\Delta
t}=0$ with probability $P_1$. The transitions are then:
\begin{equation}
\left[ \begin{array}{c}
\gamma_{i,t+\Delta t}=1 \\
\gamma_{i,t+\Delta t}=0 \\
\end{array} \right] = 
\left[ \begin{array}{cc}
1-p_1 	& 0	\\
 p_1		& 1		\\
\end{array} \right]
\left[ \begin{array}{c}
\gamma_{i,t}=1 \\
\gamma_{i,t}=0\\
\end{array} \right] .
\end{equation}
Any cells that have stopped moving and are touching are defined to be
part of a group or colony, denoted by $C_i$.

Next, I add an interaction between cells that are not touching.
Generally, cells interact indirectly, for example by responding to
chemicals released by other cells via chemotaxis. These responses can
have the effect of moving the cells toward or away from each other on
average, as if there were some effective force between the cells. I
use a model with direct interactions in the form of forces between
cells as a proxy for the behaviour we expect to see due to chemotaxis
in response to chemicals released by other cells. This approach is
fairly common \cite{lee:2001,mogilner:2003}. Lee {\it et al.} (2001)
examined a one dimension, continuous ODE model of chemotaxis, and
showed that a non-local approximation for processes such as chemotaxis
is valid if the rate of diffusion of a chemotactic signal is much
faster than the rate of diffusion of the organisms. This is the case
with many bacteria.  Lee {\it et al.} further argue that this kind of
approach facilitates learning about general properties of a system
without needing to take into account the dynamics of faster scale
processes.  Additionally, modelling the diffusion of chemicals that are
being produced and consumed by the bacteria can increase computational
requirements, without providing significant gains in understanding. So
instead of modelling chemotaxis explicitly, I assume that the cells in
this model interact via non-local or ``direct'' interactions:
\begin{eqnarray*}
d{\bf X}_i &=& \gamma_{i,t} \phi_i
		{\bf V}_i dt  \\ 
d{\bf V}_i &=& \frac{1}{m}F_{i}dt + q d{\bf W}.\\
\end{eqnarray*}
Here $F_i$ is the force on the $i$th cell due to all other cells,  
\begin{equation}
F_i =  \sum_{\stackrel{i=0}{ i \ne j}}^N f_{ij}(|{\bf X}_i - {\bf X}_j|), \label{eq:tot_force}
\end{equation}
where $f_{ij}$ is the force on $i$ due to $j$, which depends only on
the distance between the cells. This force can be attractive or
repulsive, and differs in strength depending upon the functional form
chosen for $f_{ij}$. There are various possible functional forms of
non-local forces that have been used to model interactions between
organisms. Many of these options are gradient type forces
\cite{mogilner:2003}. For example, Lee {\it et al.} (2001) showed that
in one dimension an ``effective interaction force'' for chemotactic
cells can take the form of a decaying exponential. Also typical are
inverse power forces, with the form \cite{mogilner:2003}:
\begin{equation*}
F(x) = \frac{A}{x^n} - \frac{R}{x^m}.
\end{equation*}
I use a force of this type for the interaction between cells in this
model. The value of the power law used is influenced by geometric
considerations. The concentration of a chemical dispersing from a
point source in three dimensions would fall off
as the square of the distance to the source. I also expect that a cell
will be influenced more by nearby cells than by distant cells. This is
analogous to a physical force like a gravitational force. I therefore
choose a functional form for the effective force between the cells
that is inversely proportional to the square of the distance between
cells:
\begin{equation}
f_{ij} = \frac{A_{i,j}}{{\bf R}_{ij}^2} {\bf \hat{R}}_{ij}, \label{eq:pair_force}
\end{equation} 
where $A_{i,j}$ is the force constant between cell $i$ and $j$ and
${\bf R}_{ij} = {\bf X}_i - {\bf X}_j$ is the vector between cells $i$
and $j$. The unit vector from $i$ to $j$ is denoted by ${\bf
\hat{R}}_{ij}$. If $A_{ij} > 0$ the force is repulsive, and if
$A_{ij}<0$ it is attractive.

This model can also be generalised to include attraction toward, or
repulsion from, fixed sources of chemical signals. For instance,
imagine there was an external point source of a chemical located at
${\bf X_{s}}$, and the effective force between cell $i$ and this
source, which I will denote as $F_{ext}$, that has a form similar to
Eqn. \ref{eq:pair_force}
\begin{equation*}
F_{ext} = \frac{B}{|{\bf X}_i - {\bf X}_s|^2} {\bf \hat{R}}_{is}
\end{equation*}
where $B$ is the force constant and ${\bf \hat{R}}_{is}$ is the unit
vector from cell $i$ to the chemical source. Then the total force on
the $i^{th}$ cell (Eqn. \ref{eq:tot_force}) would become:
\begin{equation*}
F_i =  F_{ext} + \sum_{\stackrel{i=0}{ i \ne j}}^N f_{ij}(|{\bf X}_i - {\bf X}_j|). 
\end{equation*}
This is similar to looking at particles undergoing Brownian motion in
a field of force \cite{kramers:1940}

\subsection{Cell death and reproduction}
The model thus far is only useful over fairly short time periods,
{\it i.e.} periods much less than the average doubling time of a cell, so
that the size of the population is unlikely to change. For periods
longer than this, it is important to include population dynamics, in
the form of births and deaths. In order to do this I use four
more state variables: $\alpha_i$, $T_i$, $a_i$, and $\tau_i$ (see
Table \ref{tb:statevars1}). The first, $\alpha_i$, indicates whether or
not the $i^{th}$ cell is alive. Cells transition from $\alpha_{i, t}=1
\rightarrow \alpha_{i,t+\Delta t}=0$ with probability $P_2$:
\begin{equation*}
p(\alpha_{i,t+\Delta t}=0|\alpha_{i,t}=1)=P_2.
\end{equation*}
If $\alpha_i=0$, then $\gamma_i=0$, {\it i.e.} the cell also stops
moving. Additionally, a living cell can double in a doubling period
$T_i$. The number of times a cell has divided is the cellular age,
$a_i$. Finally, the number of time steps since the last doubling of a
cell will be denoted by $\tau_i$. At every time step, $\tau_i$ is
incremented. When the time since the last doubling, $\tau_i$, reaches
the doubling time, $T_i$, {\it i.e.}, $\tau_i=T_i$, the cell doubles,
the cellular age $a_i$ is incremented, and the time since the last
double is reset, so $\tau_i=0$. With these population dynamics
included, the state of the $i^{th}$ cell at time $t$ is given by:
\begin{equation*}
\mb{S}_i(t) = \{ \mb{X}, \mb{V}, s_{i,t}, T_i, \tau_i,  C_i, r_i=r \}_t
\end{equation*}
where $s_{i,t}$ is the stage state of the cell described in Section \ref{cellstate}.

The equations for the change in position and velocity over time need
to be modified slightly, in order to introduce deaths into the system:

\begin{eqnarray}
d{\bf X}_i &=& \alpha_{i,t} \gamma_{i,t} \phi_i
		{\bf V}_i dt \\ c
d{\bf V}_i &=& \frac{1}{m}F_{i}dt + q d{\bf W}.\\
\end{eqnarray}

\subsection{Re-spacing cells within a colony}
Finally, we need to allow for interactions between individuals within
a particular group or colony, $C_i$. Since more than one physical cell
cannot occupy the same space at the same time, it is important to
include at least a simple mechanism to space cells out as they
reproduce within a colony.

This spacing could be approached in various ways \cite{picioreanu}.
Explicitly balancing forces between cells in a colony would be
computationally expensive. Placing new cells at the perimeter of the
colony is an alternative, but later exploration of phenotypic or
genotypic variation within a colony using this framework would be
impossible. Instead, I use a simple movement heuristic where
overlapping cells will move away from each other, {\it i.e.}, cells
shove each other out of the way. This method is similar to that used
in BacSim \cite{kreft:2001}. The algorithm proceeds as follows:

\begin{enumerate}
\item Check if the $i$th cell overlaps with any other cells, {\it
    i.e.} if $\phi_i = 0$.
\item Calculate the unit vector from the $i$th cell to each of the
  overlapping cells.
\item Sum these vectors, and move the $i$th cell a distance $D$ away
  from this new direction.
\item Repeat 1-3 over all the cells in a colony.
\end{enumerate}

The distance to move the cell, $D$, will influence how quickly the
colony re-adjusts itself. In general, if this distance is more than
the distance required to move the $i$th cell away from its nearest
neighbour, the colony will become disjoint. The choice of $D$ will
therefore determine the density and compactness of the
microcolonies. If reproduction stopped, the microcolonies would end up
arranged such that cells would be just touching, as long as $D$ is
less than or equal to the minimum overlap distance between
cells. Also, if a cell from one colony overlaps with a cell from
another colony at any time, these two colonies merge to form one
larger colony.

\section{Model Simulations} \label{modelsims:1_2}
The physical space of the simulation is a two dimensional square space
with reflecting boundaries. Cells are initially randomly distributed
in the space, and the only sources of attraction or repulsion in the
system are other cells. At each time step the simulation of the full
model proceeds as follows:
\begin{enumerate}
\item Move all of the cells that are not already stopped.
\item Stop any of the cells that have collided with other moving cells
  or with stopped cells and add them to the appropriate colony.
\item If cell $i$ overlaps with cell $j$, but they are in different
  colonies, ({\it i.e.} $C_i \ne C_j$), then merge the colonies.
\item Consider killing any living cell w.p. $P_2$.
\item Consider stopping any living, moving cell w.p. $P_1$.
\item If any remaining living cells have $t_i=T_i$, create a new cell
  a distance $2r + \epsilon$ from the dividing cell, update $a_i$ and
  reset $t_i$. Increment $t_i$ for any cells that do not divide.
\item Within each colony, re-space cells using the shoving algorithm.
\end{enumerate}

Sub-models can be formulated by excluding portions of the above
algorithm or varying parameters in order to see what the effects of
various processes have upon the distributions of cells in space.

\subsection{Model with clumping and direct interactions between cells.}\label{modelsims:1}
First I explore a sub-model where cells can clump and sense each
other, but cannot reproduce or die, and do not re-space themselves
within a colony. All simulations presented in this section consist
of 5000 cells initially distributed randomly in a 500 x 500 square
space (in units of the cell radius). All simulations began with the
same initial distribution. I also set $q=1$, and assume reflecting
boundaries.

\begin{figure}[ht!]
\begin{center}
\begin{tabular}{cc}
\includegraphics[trim = 30 0 30 50, clip=true, scale=0.405]{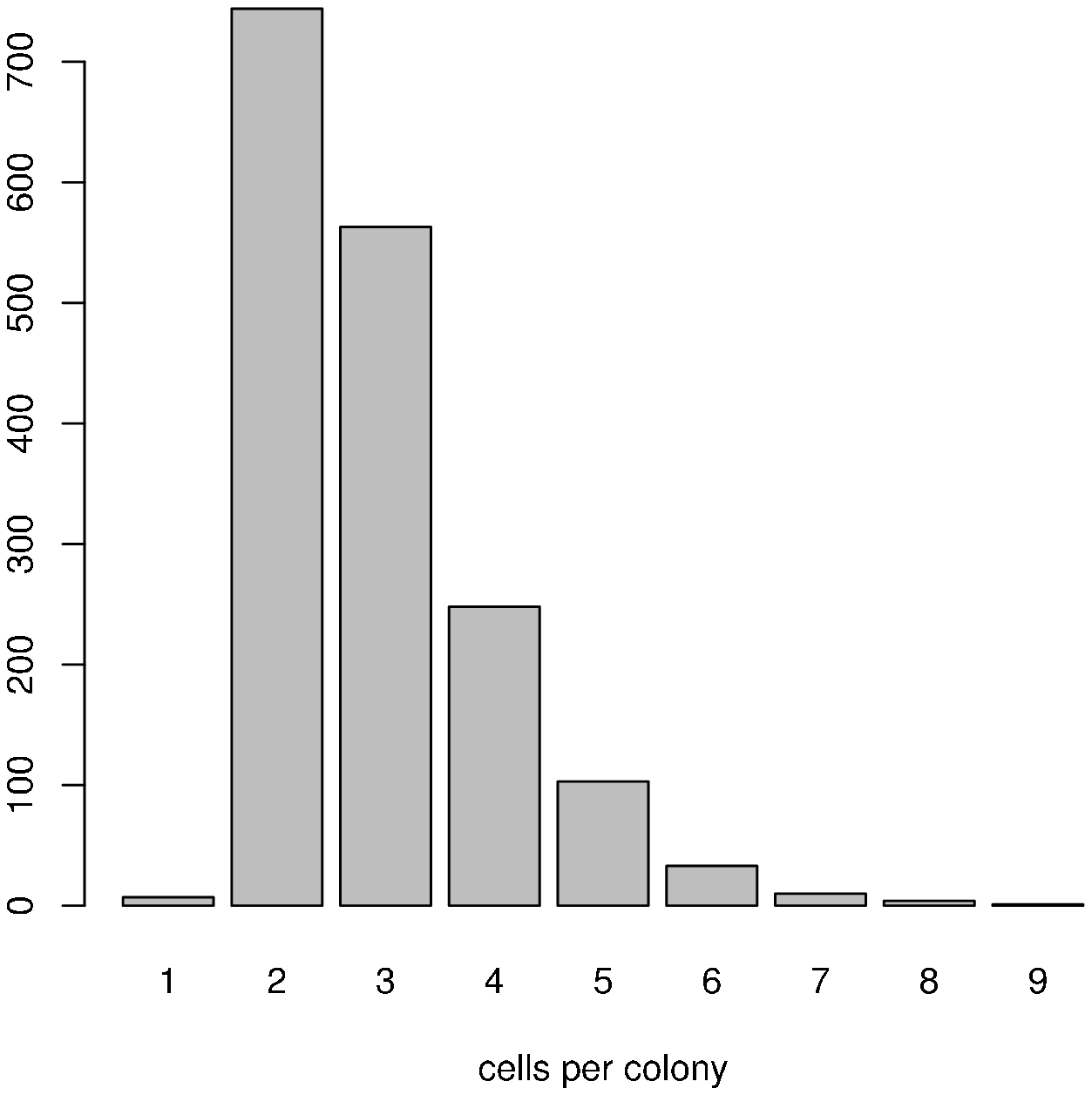} & 
\includegraphics[trim = 30 0 30 50, clip=true, scale=0.405]{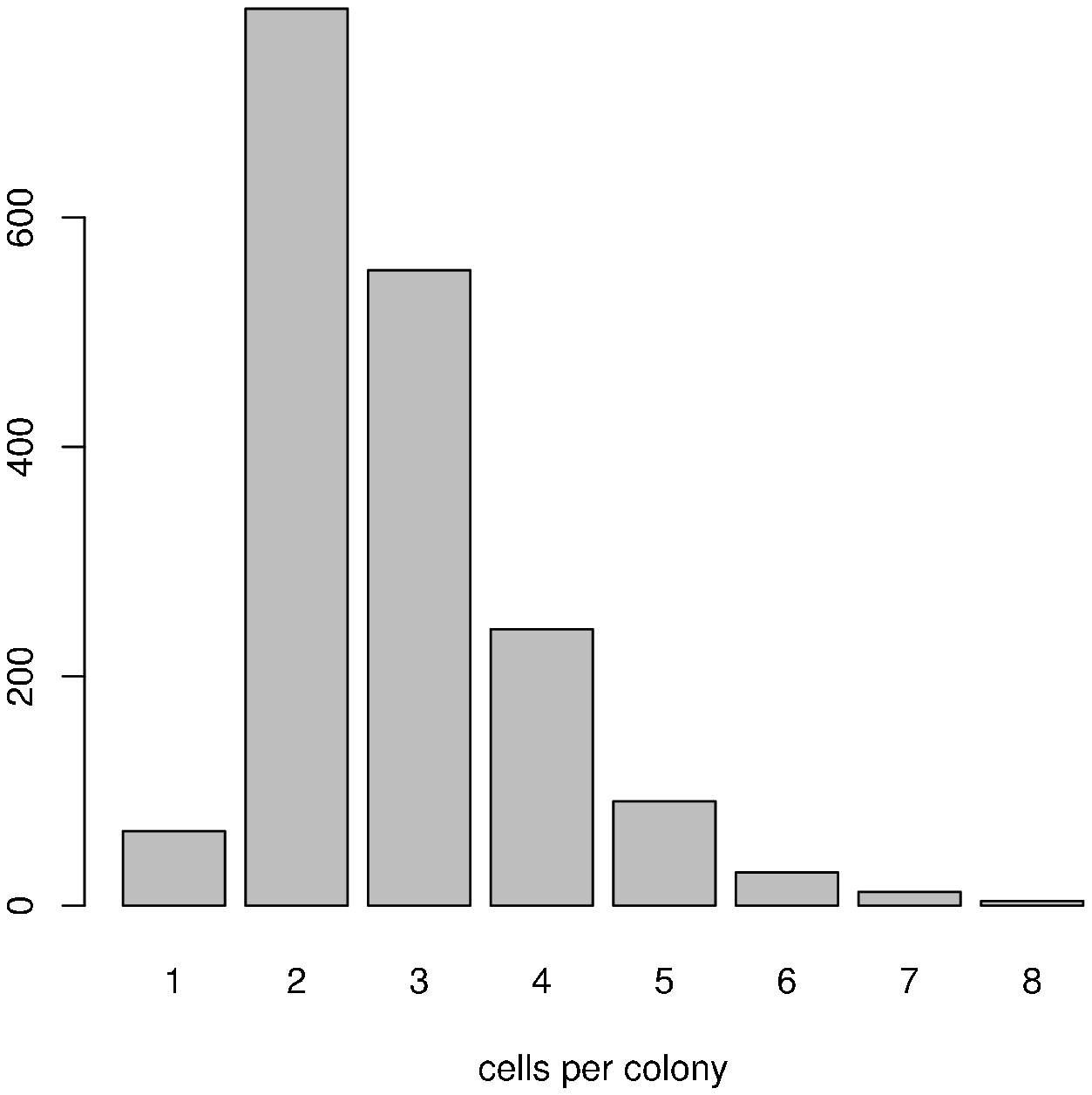} \\ 
{\small (a)} & {\small (b)} \\
\end{tabular}
\begin{tabular}{c}
\includegraphics[trim = 30 0 30 50, clip=true, scale=0.405]{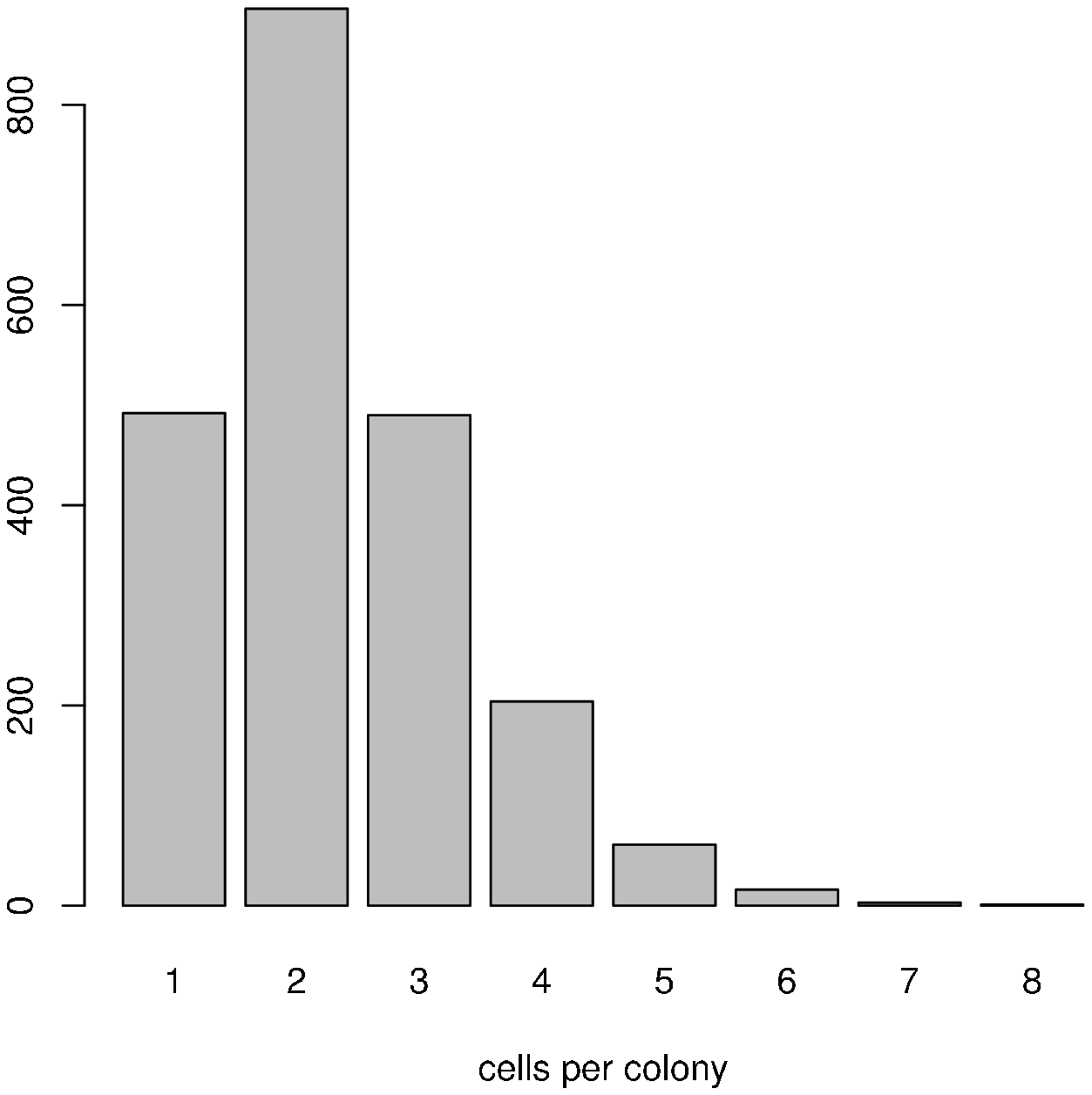} \\
{\small (c)}\\
\end{tabular}
\end{center}
\caption[Distribution of colony sizes without interactions between
cells]{Distribution of colony sizes for model with stopping and
  clumping, but no interactions between cells for three values of the
  stopping probability, $P_1$ (a) $P_1=0.0001$ (b) $P_1=0.001$ (c)
  $P_1=0.01$, after all cells have stopped moving.}
\label{fig:distrib_M1}
\end{figure}

Under this model, the distribution of colony sizes after all cells
have stopped moving for three values of the stopping probability,
$P_1$ with no forces between cells ($A_{i,j}=0$) are shown in Figure
\ref{fig:distrib_M1}. Increasing the stopping probability mainly acts
to shift more cells from colonies of size 2 into colonies of size 1
(as can be seen when comparing Figure \ref{fig:distrib_M1}a and b).
However, as $P_1$ gets much larger, the proportion of individual cells
stopping without running into other cells increases. This results in
the formation of more colonies of size 1, and a decrease in the number
of colonies of size $>2$. (Figure \ref{fig:distrib_M1}c).

\begin{figure}[htp]
\begin{center}
\includegraphics[trim = 30 20 0 10, scale=0.4]{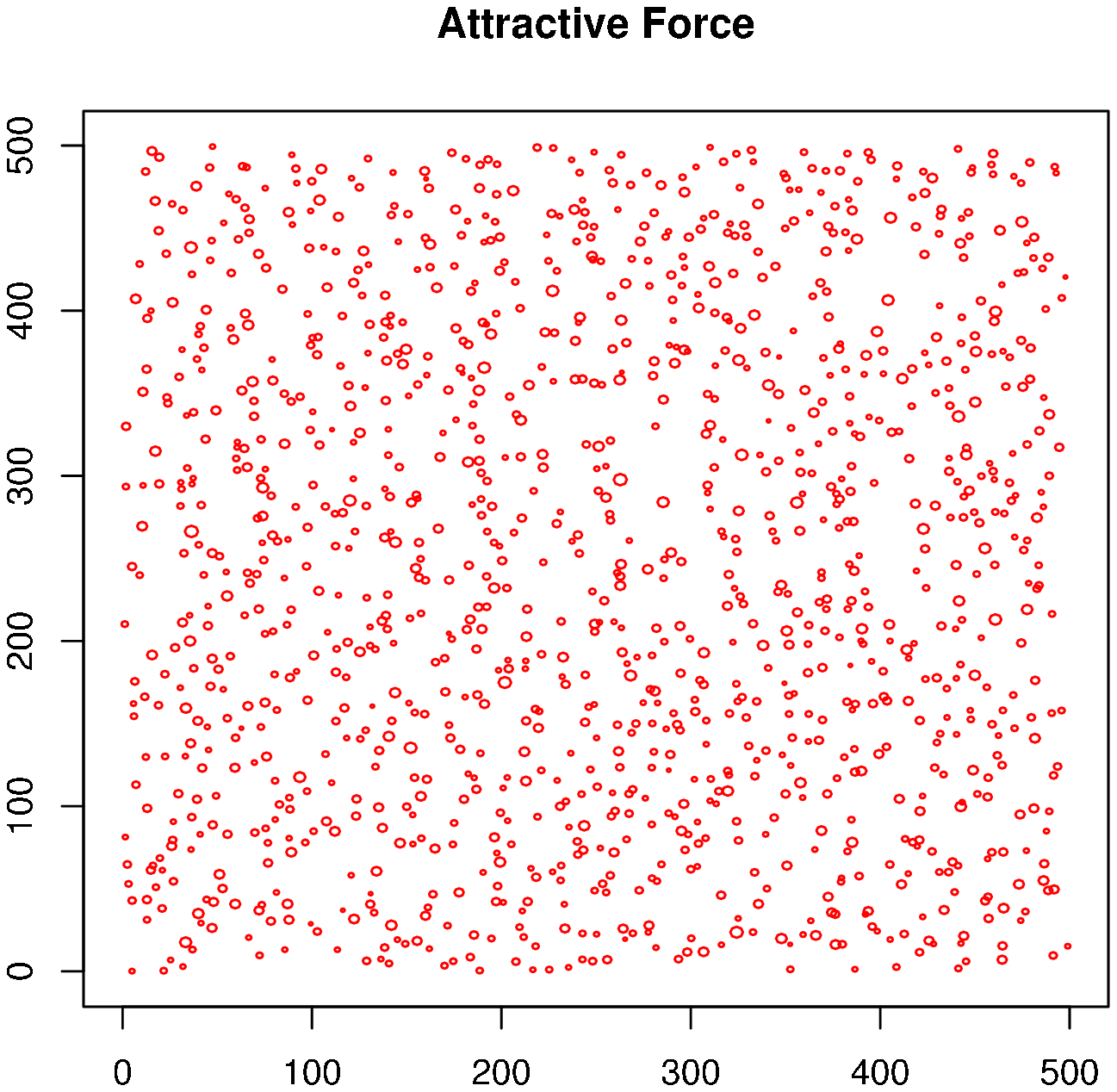}
\includegraphics[trim = 0 20 30 10, scale=0.4]{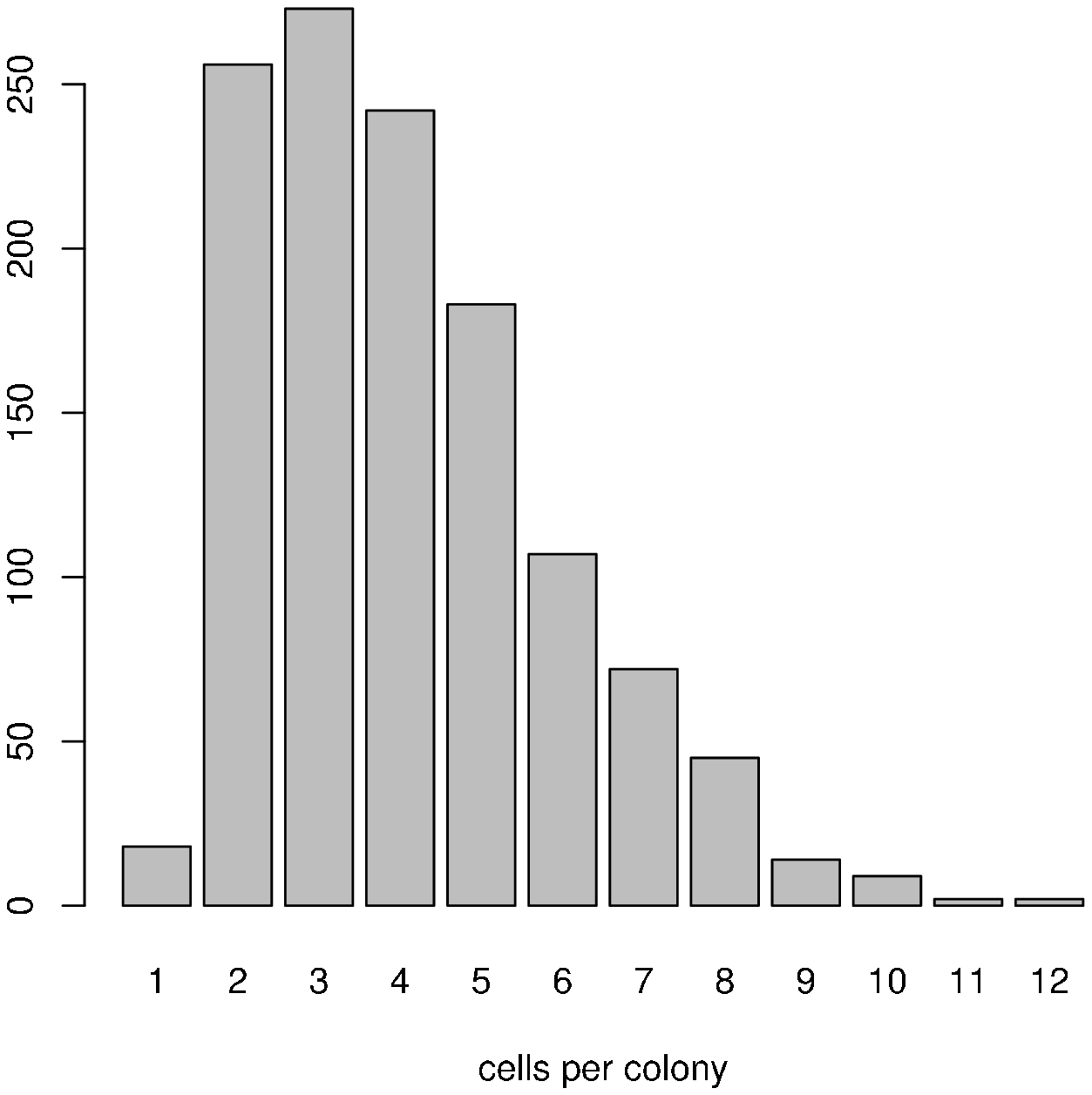} \\
\includegraphics[trim = 30 20 0 10, scale=0.4]{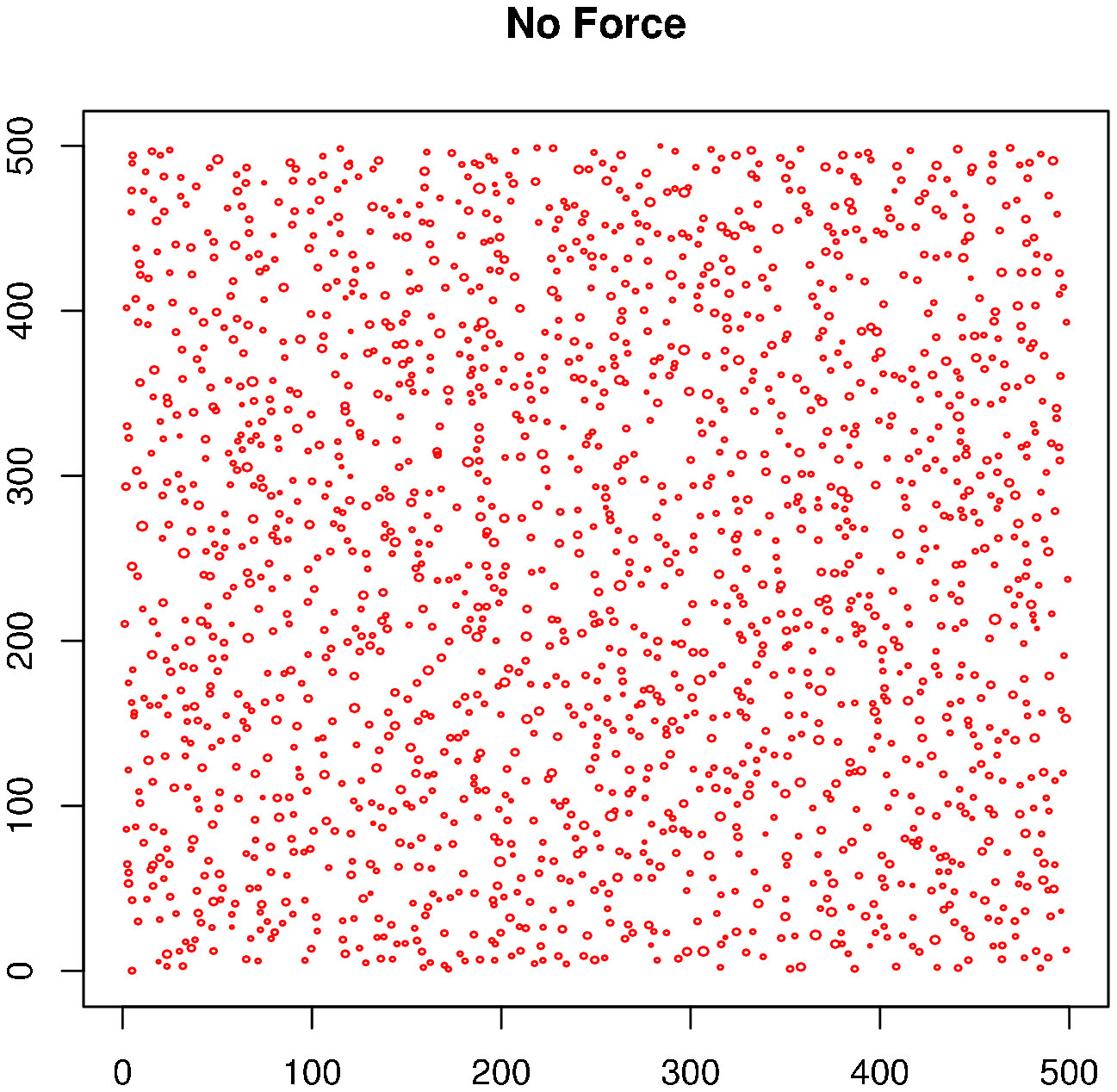} 
\includegraphics[trim = 0 20 30 10, scale=0.4]{colonies_random_001} \\ 
\includegraphics[trim = 30 20 0 10, scale=0.4]{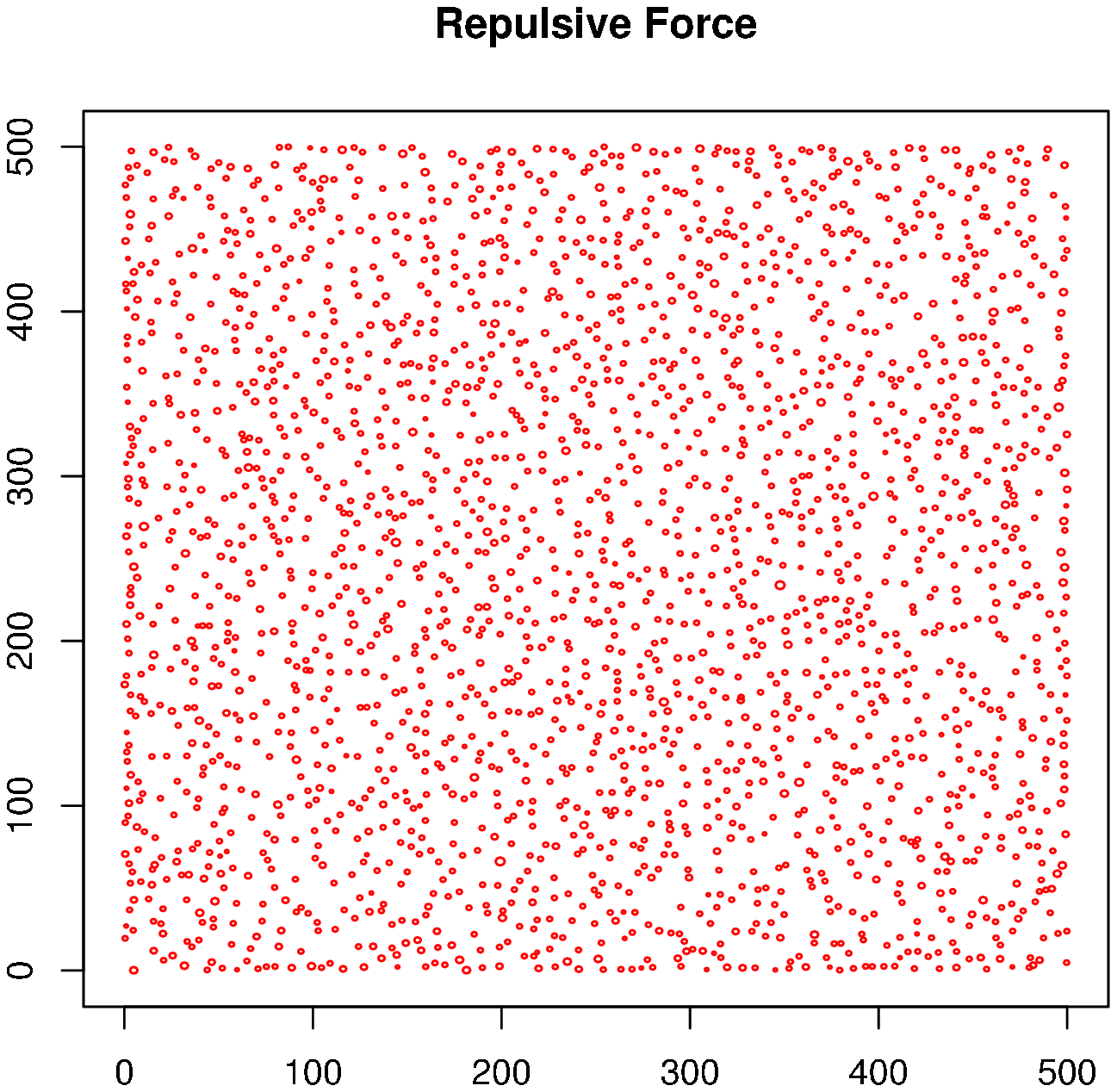} 
\includegraphics[trim = 0 20 30 10, scale=0.4]{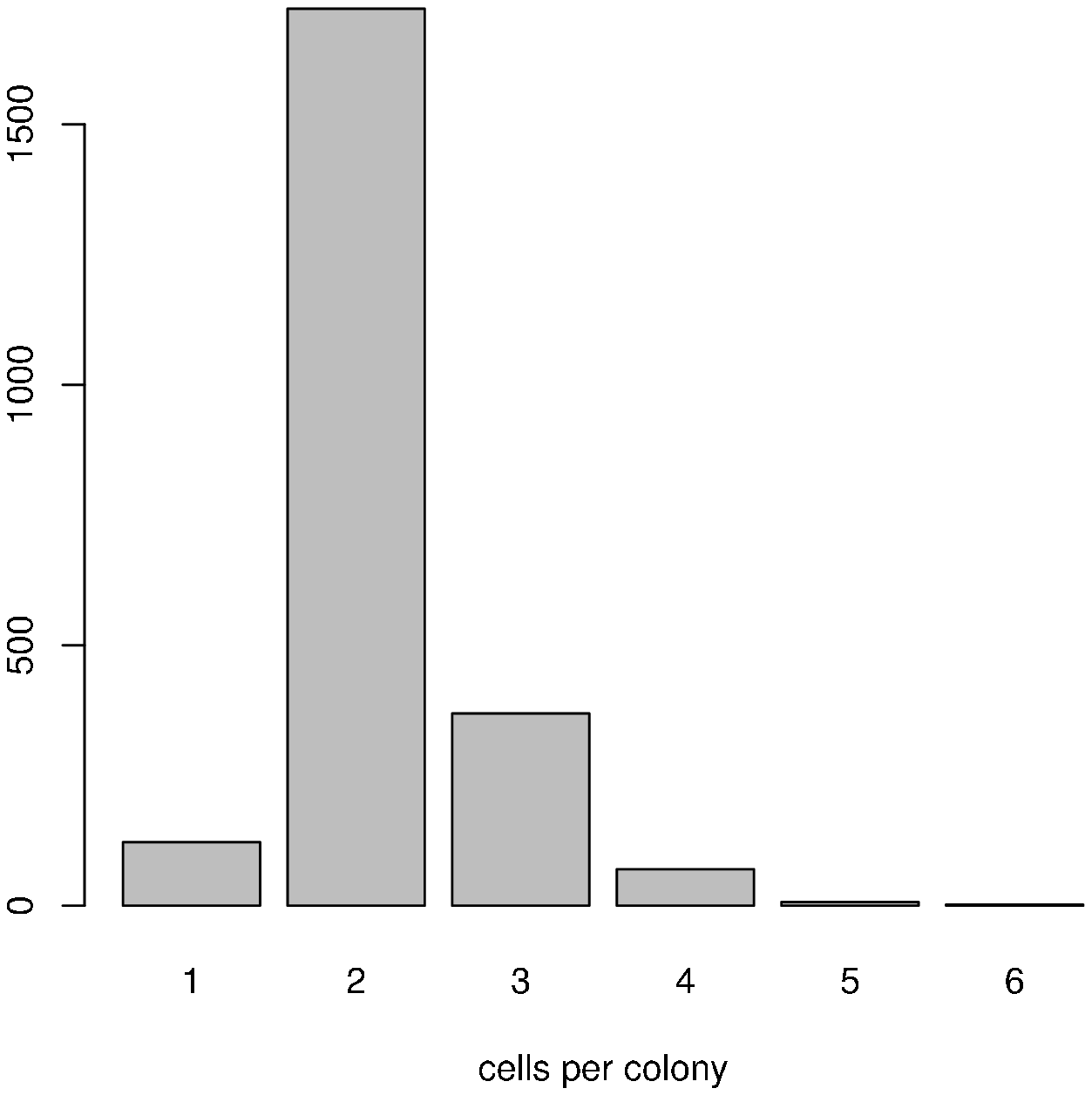} \\ 
\end{center}
\caption[Distribution of colony locations and sizes with and without
interactions between cells]{Distribution of colony locations and sizes
  with and without interactions between cells with $P_1=0.001$. (TOP)
  Attractive forces between cells; (MIDDLE) no forces between cells;
  (BOTTOM) repulsive forces between cells.}
\label{fig:distrib_M2}
\end{figure}

In Figure \ref{fig:distrib_M2} I show the final distributions colonies
and colony sizes for attractive forces, (Figure \ref{fig:distrib_M2}
a), no forces (Figure \ref{fig:distrib_M2} b), and repulsive forces
(Figure \ref{fig:distrib_M2} c). For the attractive case the force
constant $A=-10$, and for the repulsive case $A=10$.  Notice that each
of these models exhibits distinctive distributions of colony size
(numbers of cells in a colony). In particular, when there are
attractive forces between cells, the colonies are much more likely to
be large, and the distribution exhibits a longer and heavier right
hand tail.  When the cells repulse the colonies are smaller than when
they only interact randomly.

A simple $\chi^2$ test can be used to confirm that the three cases
explored here ({\it {\it i.e.}}  attractive, no force, and repulsive)
do in fact result in distinct distributions. I define the case of
motion without forces to be the ``expected'' distribution. The first
null hypothesis, $H_0$, is that the distribution from the attractive
case is the same as that of the the distribution found when there were
no forces present. The alternative hypothesis, $H_1$, is that they are
not equal. Since the $\chi^2$ test requires at least five counts in
any bin to give a good result, it is necessary to re-bin each
distribution into 7 bins: $\{ 1,2,3,4,5,6,7+\}$. This corresponds to 6
degrees of freedom in the test. A significance level of 0.001
corresponds to a $\chi^2$ value of 22.547.  For this first test, the
$chi^2$ value is $\chi^2 = 2448.036$ leading to a rejection. For the
second test, testing the hypothesis that the repulsive distribution is
the same as the no force (expected) distribution, I find $\chi^2 =
1029.437$, leading to another rejection. Thus as the 0.001 level, I
reject the hypotheses that the attractive and repulsive distributions
are the same as the random case.

These three cases can be viewed as representative of the
types of clumping that might be observed in a bacterial system. For
instance, in a system where bacteria attract via chemotaxis, we might
see distributions of microcolonies similar to those shown in Figure
\ref{fig:distrib_M2}a. However, if the chemotaxis system is disrupted,
for instance by some chemical additive or by genetic manipulation, then
the new system might look more like what is shown in Figure
\ref{fig:distrib_M2}b. If on the other hand, the system is manipulated
so that the chemotaxis system is reversed, so cells move away from
each other, the system would appear more like what we see in Figure
\ref{fig:distrib_M2}c.

\subsection{Model with clumping, direct interactions, and births and deaths}
Next I explore a sub-model that includes all portions of the full
model except shoving within a colony. All simulations presented in
this section consisted of cells initially distributed randomly in a
500 x 500 square space (in units of the cell radius). Parameter
settings for the simulations are shown in Table \ref{tb:sim_M3}.

\begin{table}
\begin{center}
\begin{tabular}{| c | c | }
\hline
Parameter & Value  \\
\hline
$N_{initial}$ & 250 \\
$t_{final}$ & 1000 \\
$P_1$ & 0.001 \\
$P_2$ & 0.00005 \\
\# replicates & 50 \\
\hline
\end{tabular} 
\caption[]{Parameter settings for simulations of model with clumping,
  direct interactions, and births and deaths \label{tb:sim_M3}}
\end{center}
\end{table}

\begin{figure}[htp]
\begin{center}
\begin{tabular}{cc}
\includegraphics[trim= 20 10 20 0, scale=0.55]{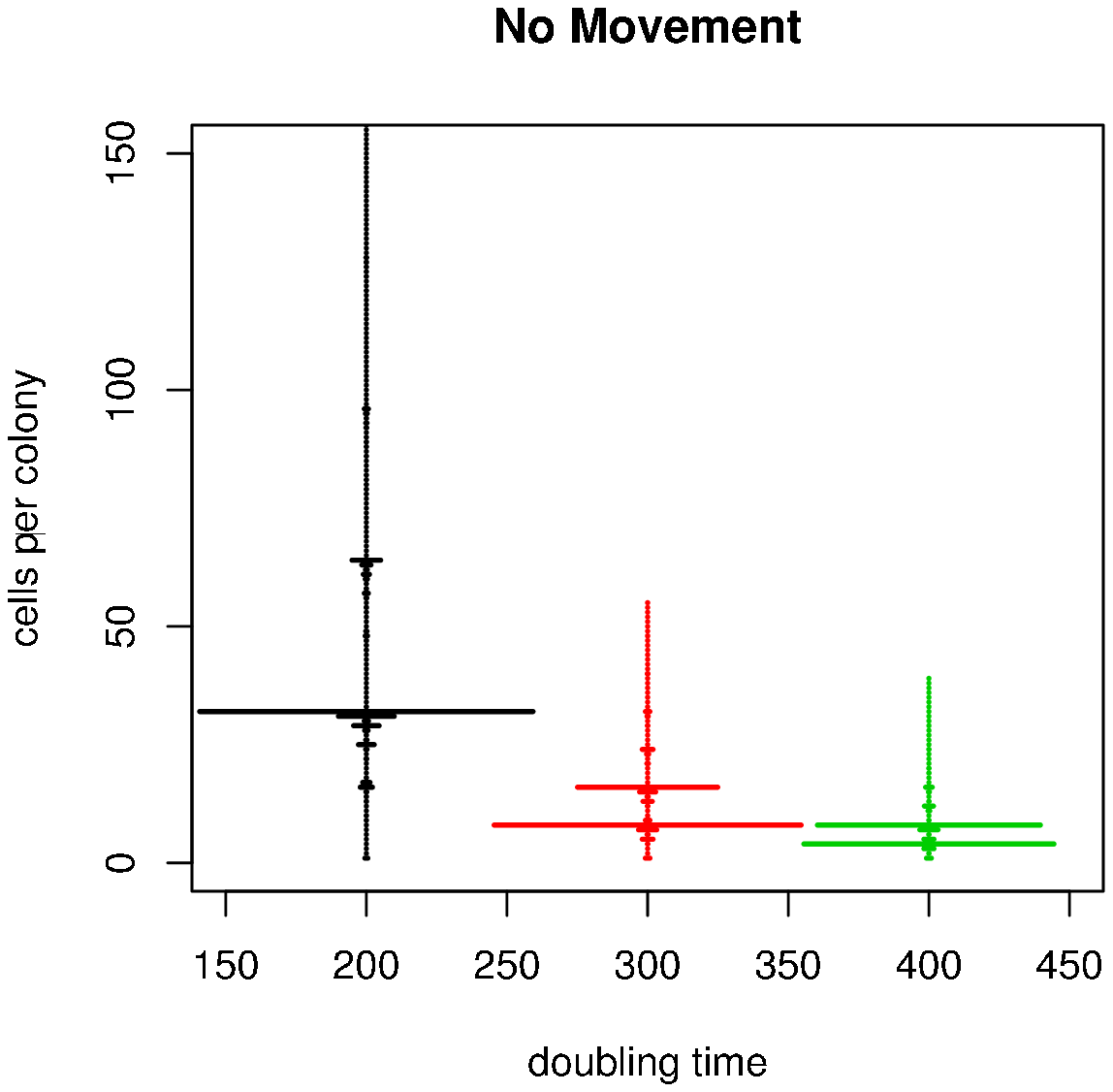} & 
\includegraphics[trim= 0 10 20 0, clip=true, scale=0.55]{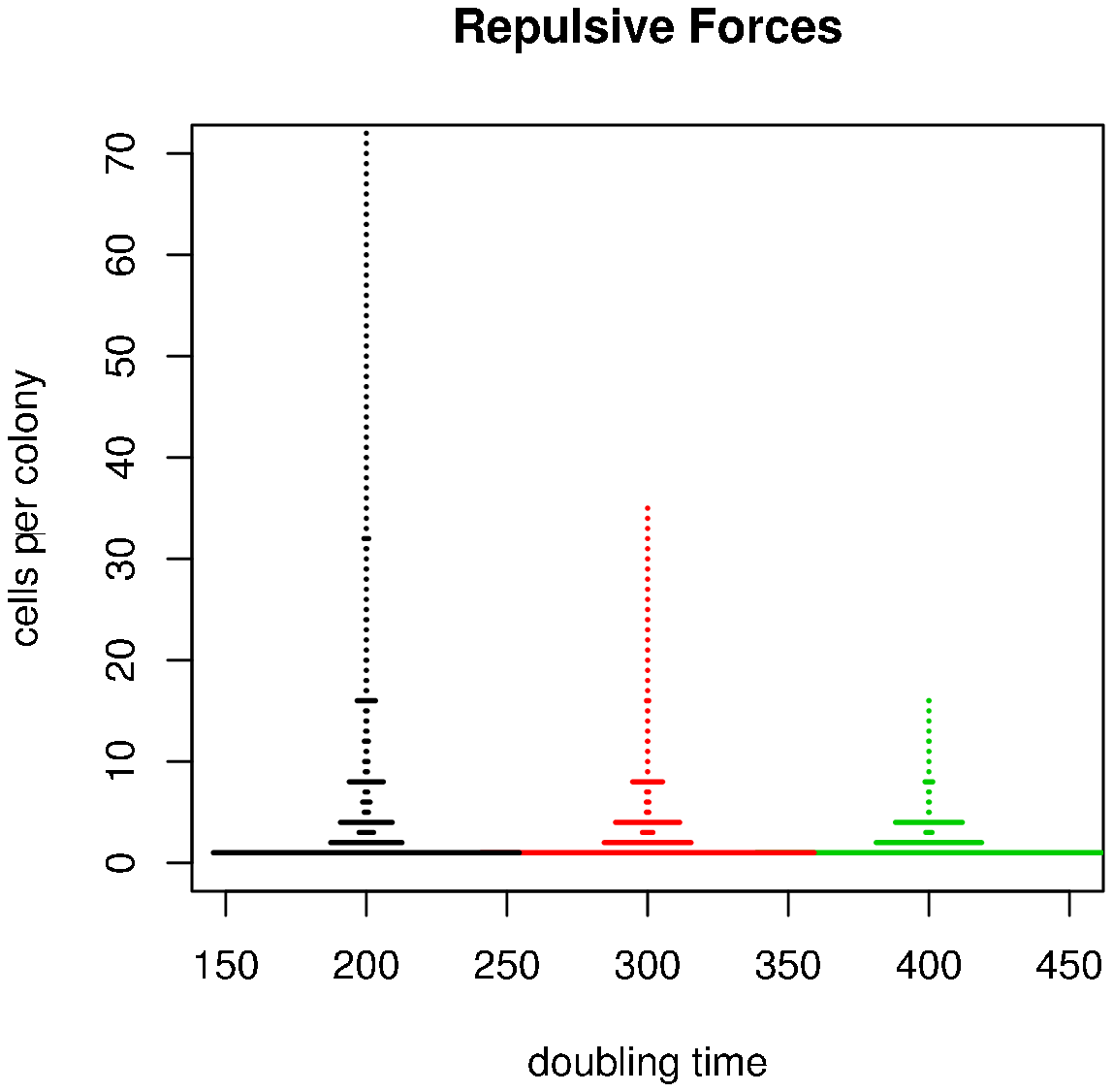}  \\
\includegraphics[trim= 20 10 20 0, scale=0.55]{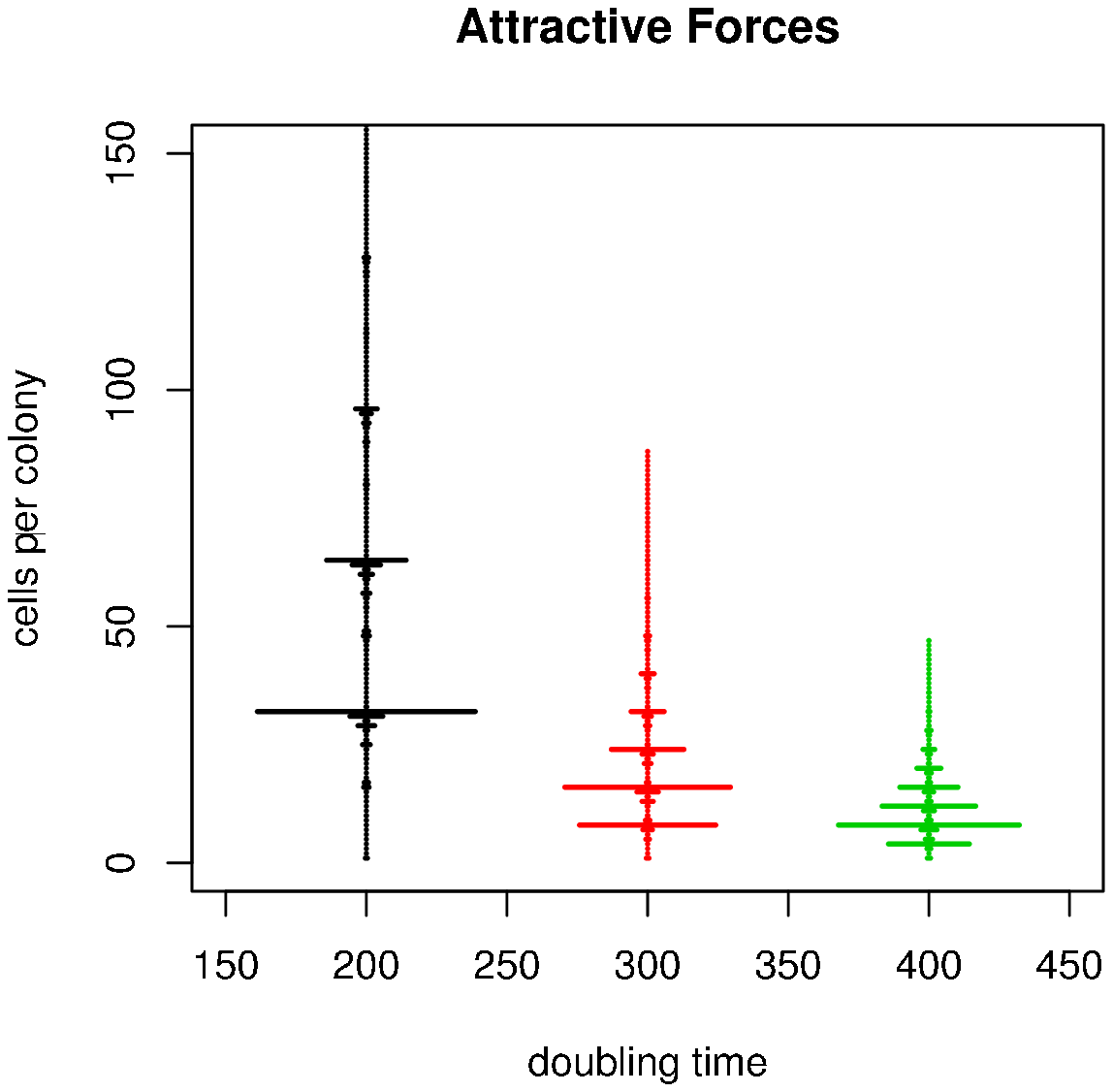} & 
\includegraphics[trim= 0 10 20 0, clip=true, scale=0.55]{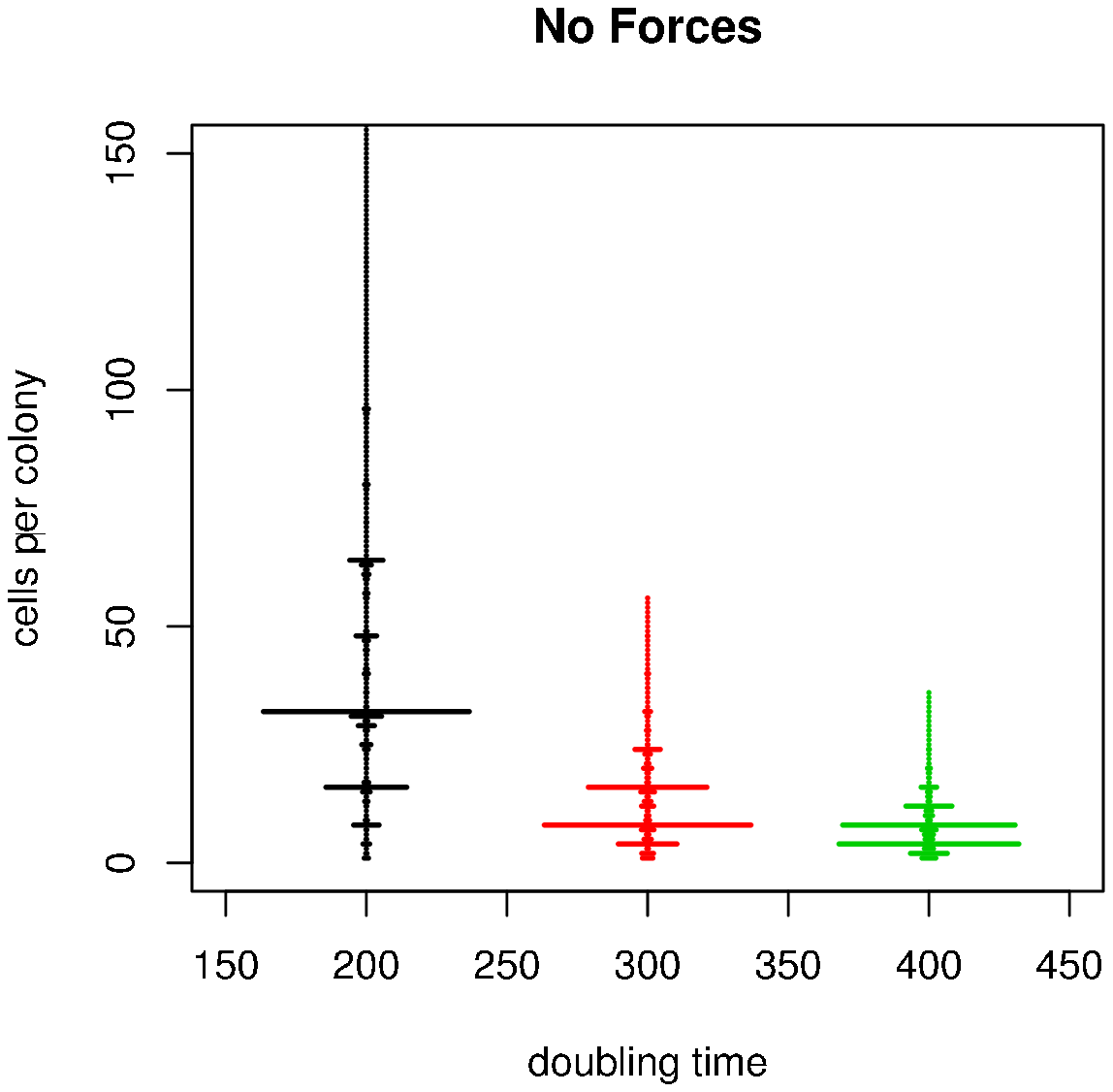}\\
\end{tabular}  
\end{center}
\caption[Alternative histograms of final colony size for model
$M_3$.]{Alternative histograms of final colony size. The width of the
  horizontal bars correspond to the counts in the histogram.}
\label{fig:sims_dots}
\end{figure}

In Figure \ref{fig:sims_dots} I show visualisations of the frequencies
of colony sizes cells aggregated over 50 replicates for 3 different
doubling times (200 steps, 300 steps or 400 steps), three force
constants ($A=-10$, $A=0$, and $A=10$ for the attractive, no force,
and repulsive cases, respectively), and a no movement case.

Except for the case of repulsive forces, where cells continue moving
for longer and form smaller colonies, the spacing between the major
peaks of the distribution corresponds to $2^d$ where $d =
\left\lfloor\frac{T_i}{t_{tot}}\right\rfloor$, where $t_{tot}$ is the
total run time, $T_i$ is the doubling time, and $\lfloor \cdot
\rfloor$ denotes the {\it floor} function, which returns the largest
integer smaller than or equal to the argument. This value of $2^d$
also usually corresponds to the mode or to the second largest peak of
the distribution. For example, for $T_i = 200$, $d=5$ so the largest
peak is at $2^5 = 32$ and other major peaks are spaced at $\sim 32$
individuals apart (e.g. in Figure \ref{fig:sims_dots} top left panel).
The other simulations show similar patterns, with variability being
caused by the interactions between individual cells, and stochastic
deaths.  There is additional variability in the simulation with $T_i =
300$ and $T_i = 400$ since these correspond to non-integer values of
$\left(\frac{T_i}{t_{tot}}\right)$. In these cases, all cells
reproduce at least $d$ times, but a proportion will reproduce $d+1$
times . However, the pattern of major peaks is approximately $2^d$
times the distribution of colony sizes for model $M_2$ (interactions
without births and deaths) as shown in Figure \ref{fig:distrib_M2}.

\subsection{Full model}

\begin{figure}
\begin{center}
\begin{tabular}{cc}
\includegraphics[scale=0.525, clip=true, trim = 30 30 20 30]{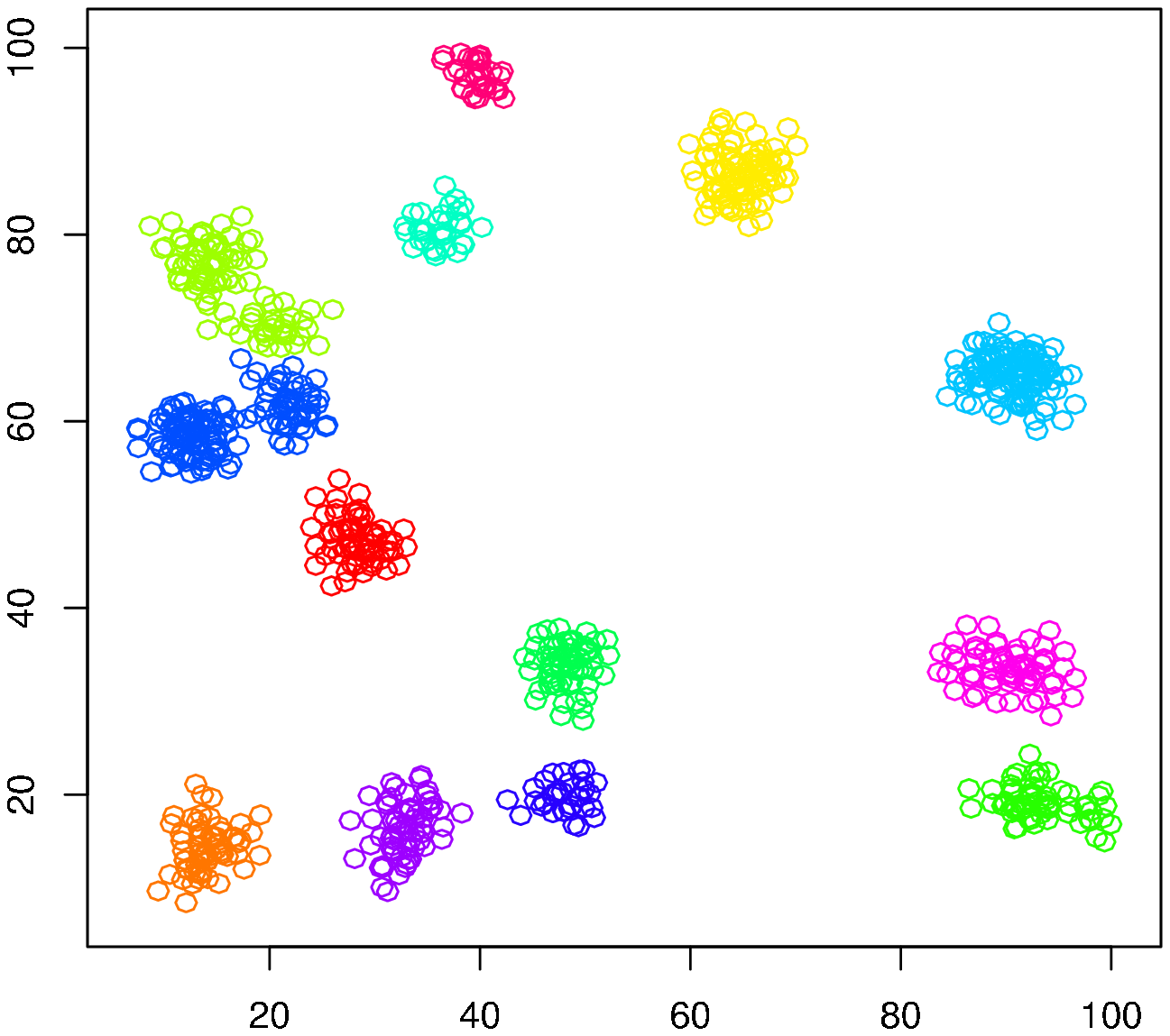} &
\includegraphics[scale=0.525, clip=true, trim = 30 30 20 30]{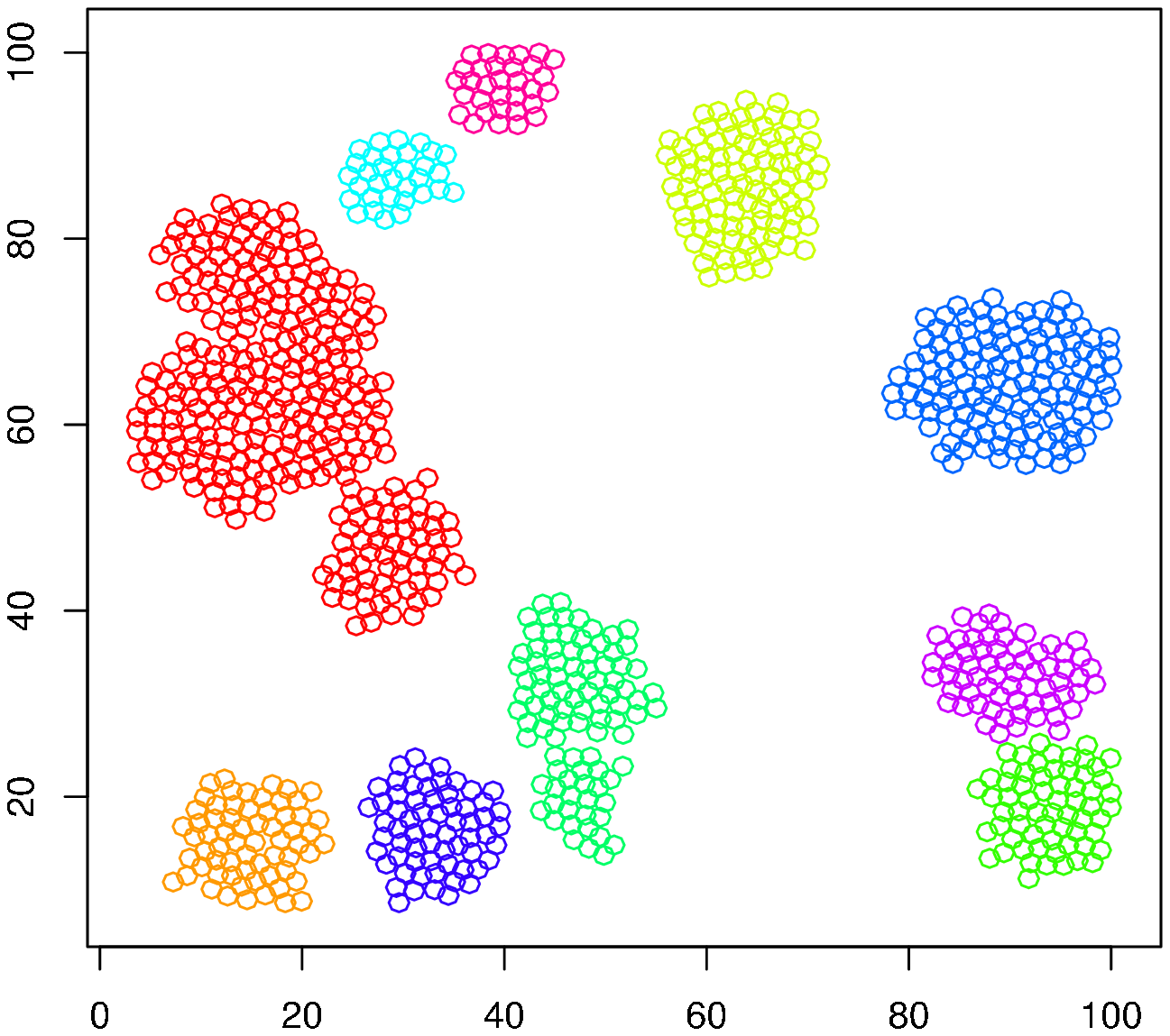} \\ 
(a) & (b)\\
\end{tabular}  
\end{center}
\caption[Cell distribution and colony membership before and after
implementation of the spreading heuristic -- attractive
example.]{Example of cell distribution and colony membership (cells in
  a colony are the same color) before and after implementation of the
  spreading heuristic. Both simulations were run starting from the
  same initial cell distribution of 30 cells. Simulation parameters:
  $T_i=200$, force=-10, $prob_{death}=0.00001$, $prob_{stop}=0.0005$,
  $T_{final} = 1000$.} \label{fig:sims_spreading1}
\end{figure}

\begin{figure}
\begin{center}
\begin{tabular}{cc}
\includegraphics[scale=0.525, clip=true, trim = 30 30 20 30]{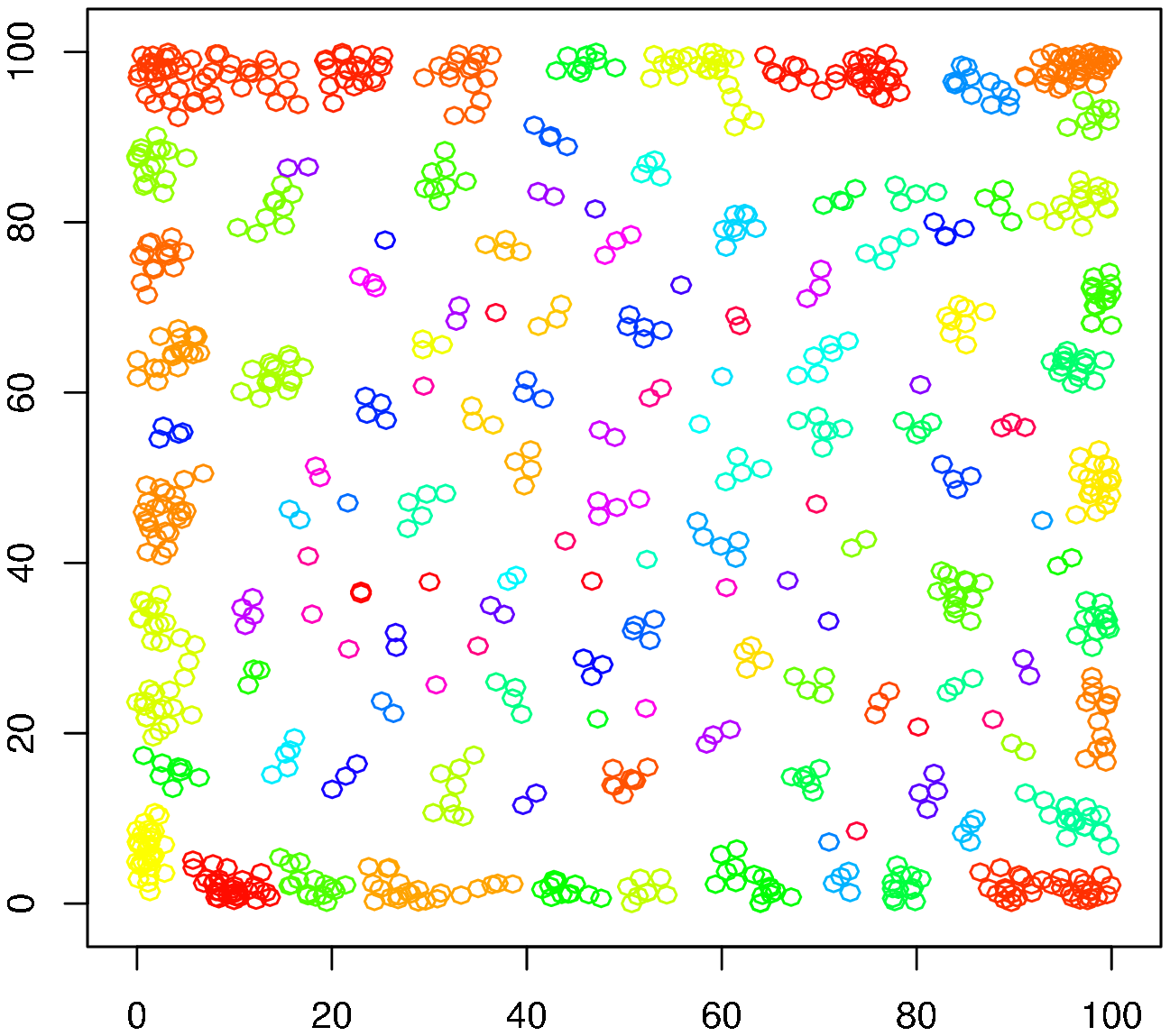} &
\includegraphics[scale=0.525, clip=true, trim = 30 30 20 30]{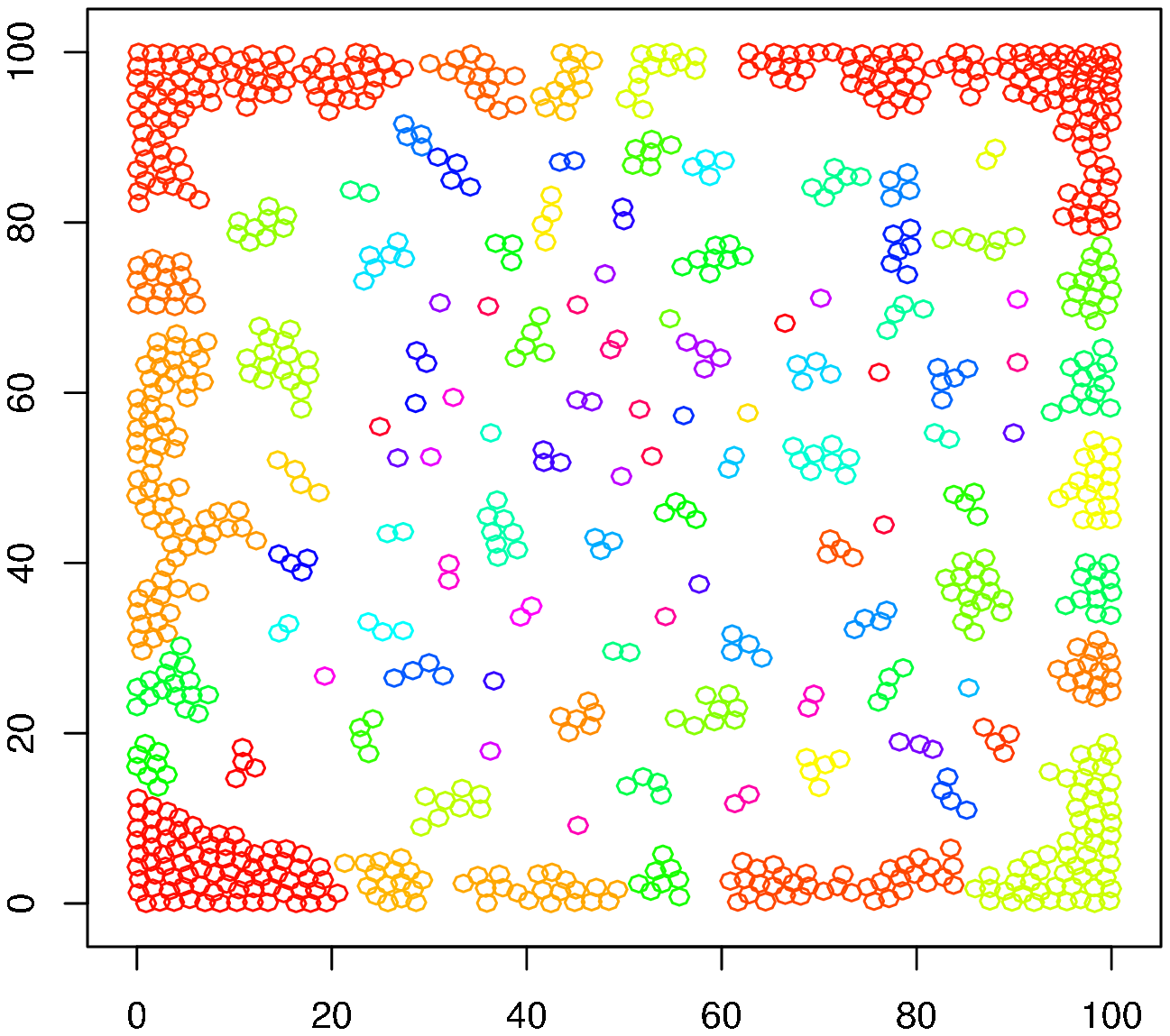} \\
(a) & (b) \\
\end{tabular}  
\end{center}
\caption[Cell distribution and colony membership before and after
implementation of the spreading heuristic -- repulsive
example.]{Example of cell distribution and colony membership (cells in
  a colony are the same color) before and after implementation of the
  spreading heuristic, but with repulsive forces between the cells.
  Both simulations were run starting from the same initial cell
  distribution of 30 cells. Simulation parameters: $T_i=200$,
  force=10, $prob_{death}=0.00001$, $prob_{stop}=0.0005$, $T_{final} =
  1000$. } \label{fig:sims_spreading2}
\end{figure}

Finally, I examine the behaviour of the full model. Examples of cell 
distributions on the surface before and after
implementing the shoving heuristic appear in Figures
\ref{fig:sims_spreading1} and \ref{fig:sims_spreading2}. For the
four simulations generating these distributions, the value of the
shoving distance $D$ was set to $\frac{1}{4}$ of the distance between
the $i$th cell and its nearest neighbour. We can see that colonies are
more likely to merge when their members spread themselves out. This
should increase the proportion of large colonies that are
observed. For cases where the cell density is low, we can expect this
effect to be small. However for larger populations, this effect could
be significant. For the repulsive case, the spreading also increases
the magnitude of the edge effects when the area of interest is small
or cell density is high (Figure \ref{fig:sims_spreading2}).

\begin{figure}[htp]
\begin{center}
\begin{tabular}{cc}
\includegraphics[trim= 20 10 20 0, scale=0.55]{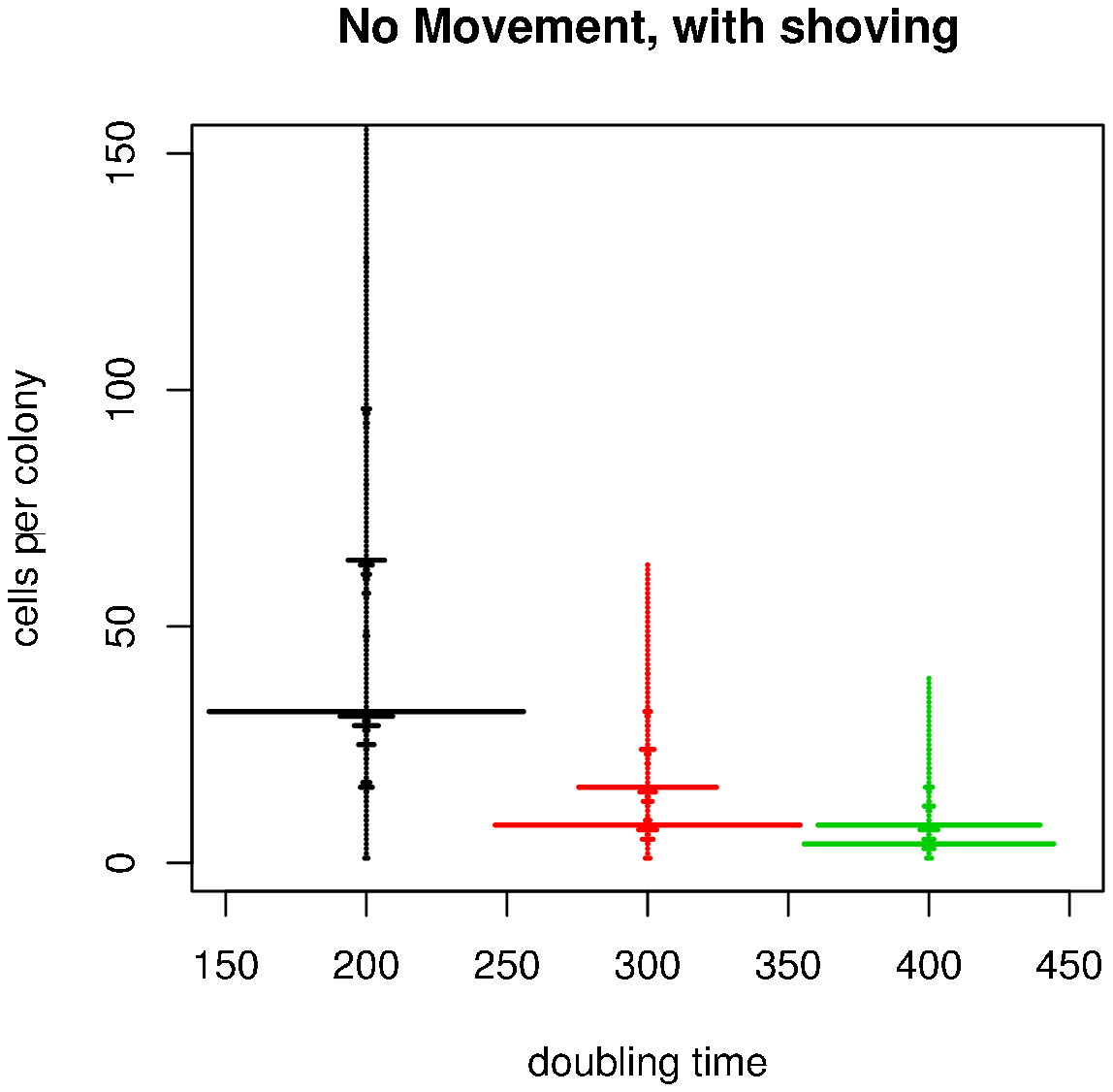} & 
\includegraphics[trim= 0 10 20 0, clip=true, scale=0.55]{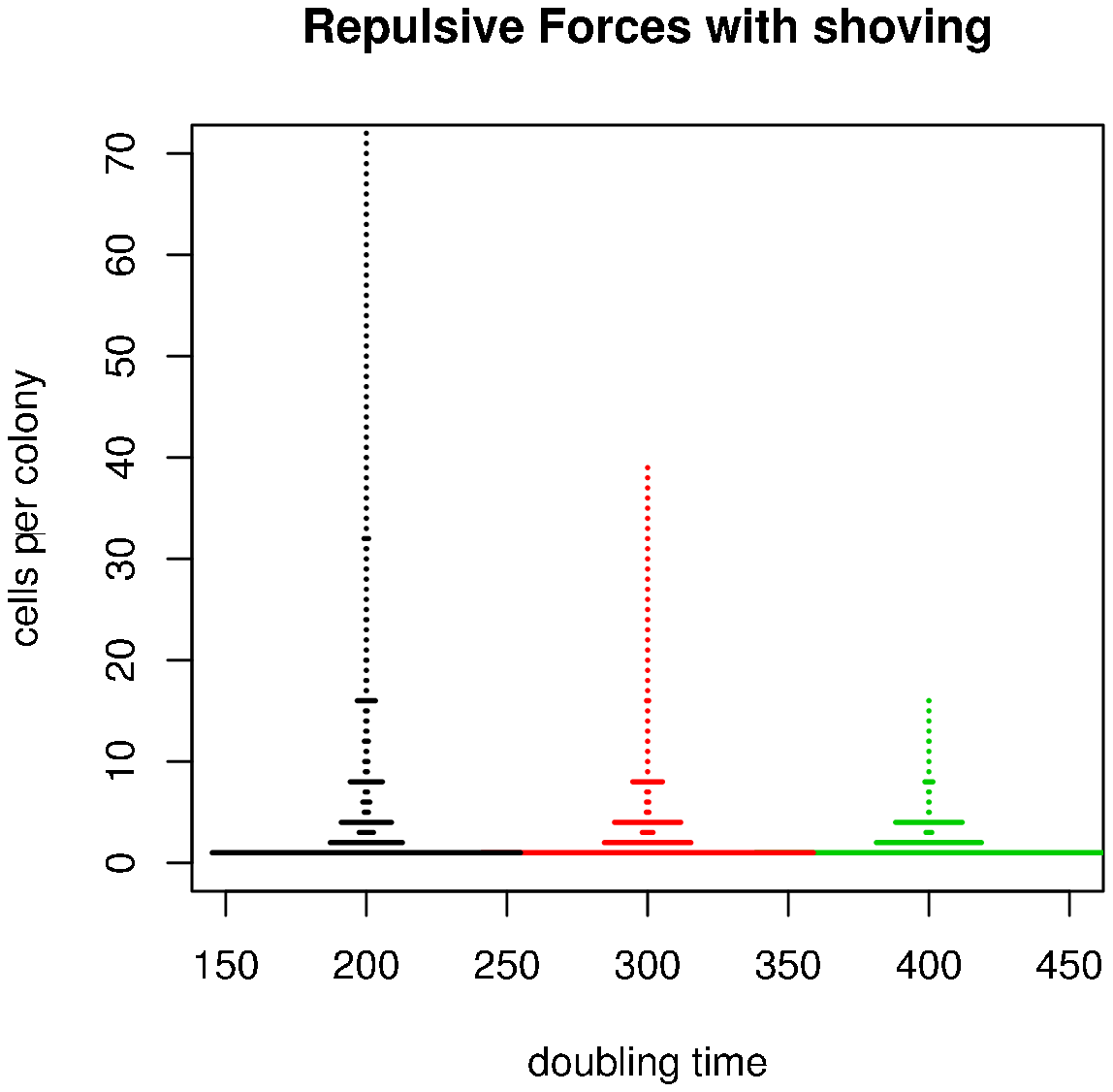}\\
\includegraphics[trim= 20 10 20 0, scale=0.55]{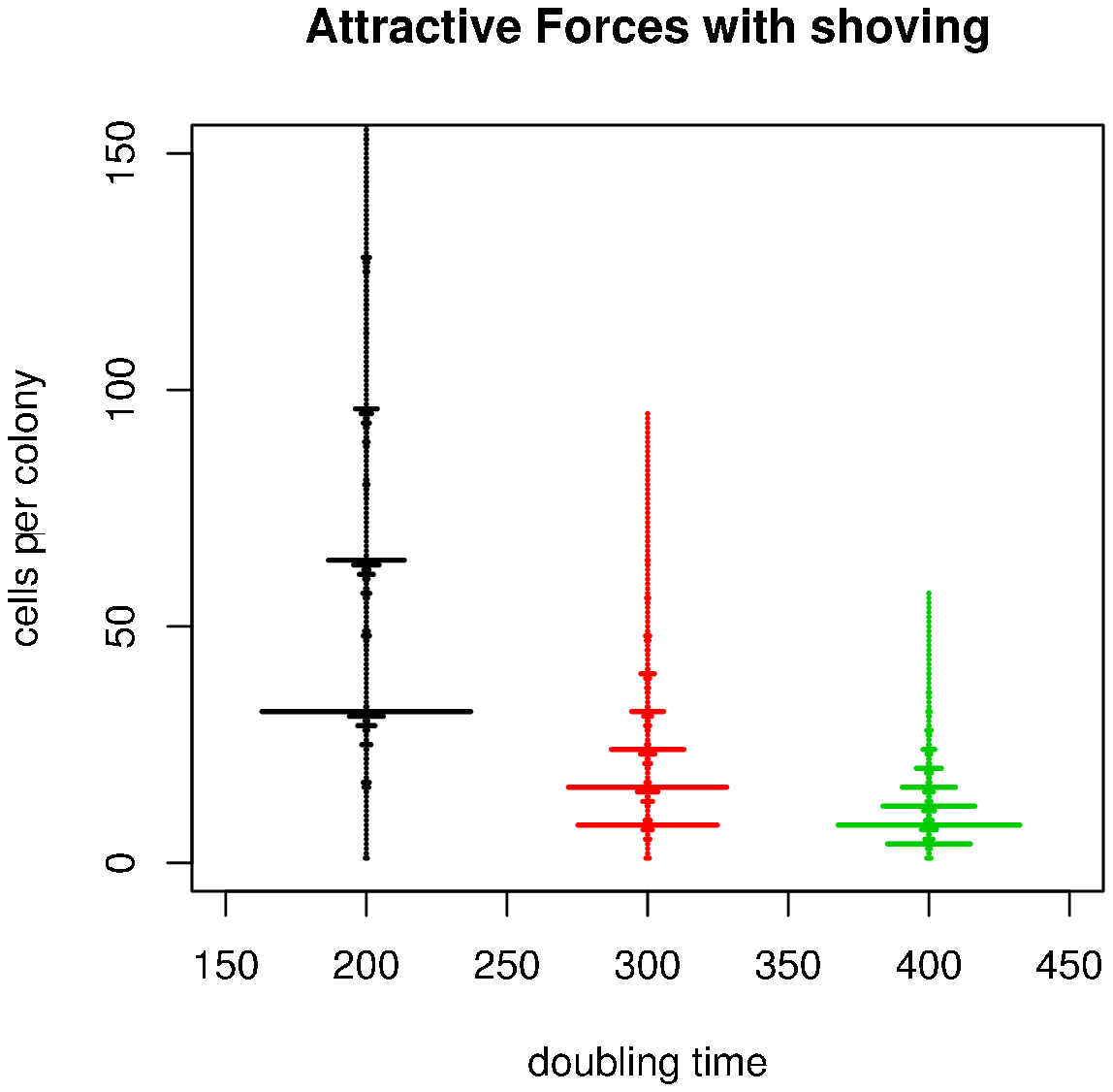} & 
\includegraphics[trim= 0 10 20 0, clip=true, scale=0.55]{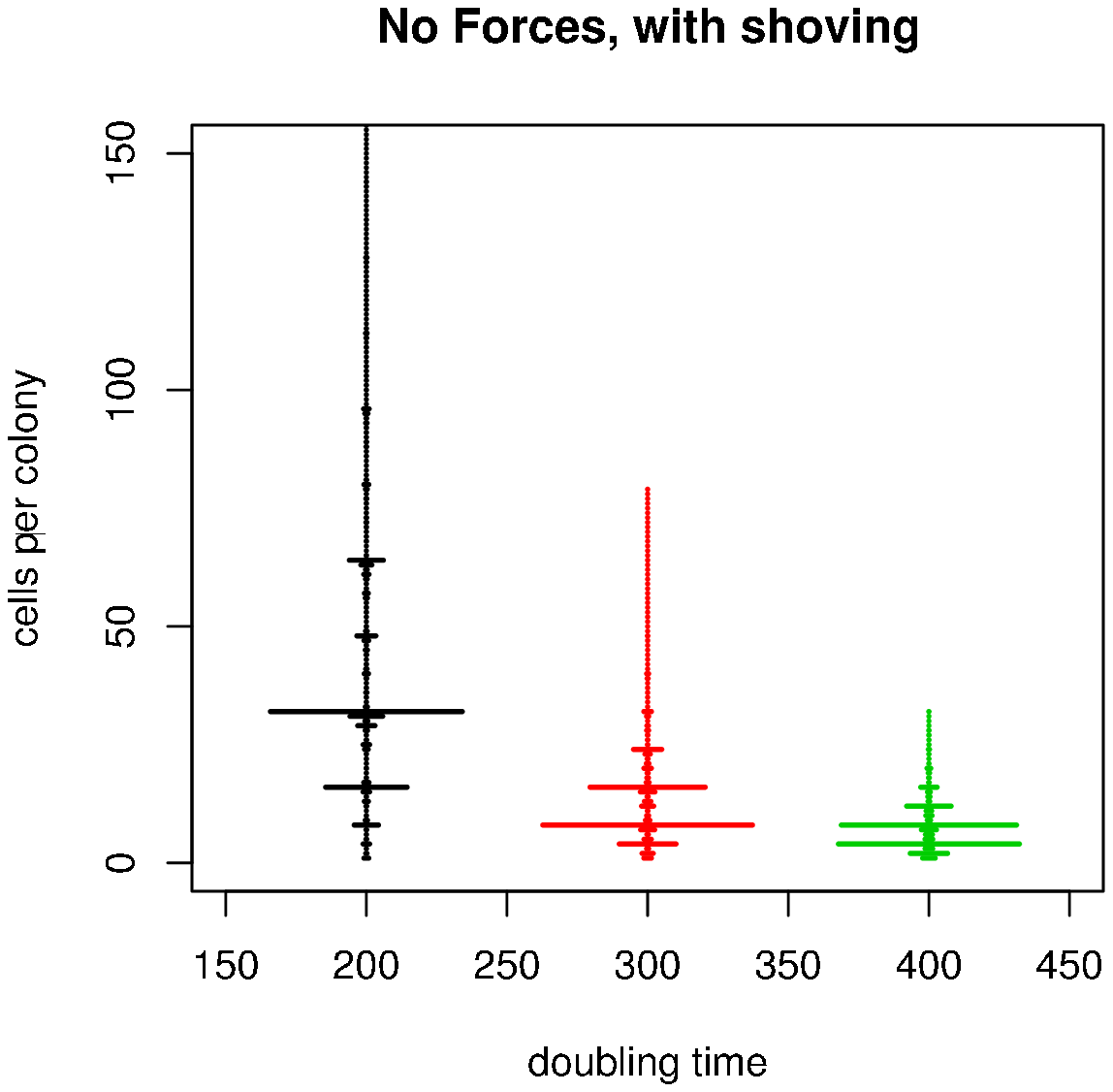}\\
\end{tabular}  
\end{center}
\caption[Histogram of final colony sizes for full model]{Histogram
  of final colony sizes for full model. Parameter settings are the
  same as for the simulations for model without shoving.  }
\label{fig:sims_spreading3}
\end{figure}

In order to explore the effect of the spreading on the observed
distributions of colony sizes, I ran simulations using the same
parameter values as in the previous section (shown in Table
\ref{tb:sim_M3}). In Figure \ref{fig:sims_spreading3}, I show plots of
the distribution of final colony sizes aggregated over 50 runs. Notice
that these distributions are very similar to those without shoving
(Figure \ref{fig:sims_dots}) except that the tails are slightly longer
and heavier. For long doubling time ({\it i.e.} $T=400$) the effect is
minimal, since the final population size is smaller.

\section{Discriminating between models}\label{model_discrim}
In order for this model formulation to be useful in approaching data,
it must be possible to discriminate between distributions of colony
sizes with different parameter settings. In Section \ref{modelsims:1}
I approached this issue by using a $\chi^2$ hypothesis test to check
if two distributions are the same. This method cannot be used for
discriminating between the distributions obtained for models that
include reproduction because the number of counts of colony sizes is
too low (and often zero) in many bins, rendering the $\chi^2$ test is
invalid. The data also cannot be re-binned without loss of information
about the structure of the distribution. Instead, a different measure
of the distance between the histograms is necessary.

\subsection{Jackknife hypothesis test}
Let $H_i$ for $i= 1, \dots, n$ be histograms summarising $n$
simulations from the model with a fixed set of parameters. Aggregating
these $n$ histograms into a single histogram, denoted by $H_a$, gives
an approximation of the target distribution of groups sizes. Imagine
that some new ``data'' histogram, $H_{data}$, is observed. We would
like to perform an hypothesis test to see if $H_{data}$ is from the
same model that generated the $H_i$'s. To do this we need a test
statistic and a way to calculate the sampling distribution of this
statistic. A simple choice for a test statistic is to calculate the
Minknowski Distance between the data and the aggregate histograms.
The $L_p$ Minknowski distance between $H_1(m; T)$ and $H_2(m; T)$,
where $m$ is the number of bins in the histogram and $T$ is the total
number of counts, is given by
\begin{equation}
D(H_1, H_2) = \left( \sum_{x=1}^n \left| H_1(x;T) - H_2(x;T)\right|^p \right)^{(1/p)}.
\end{equation}
When $p=2$, this distance gives the Euclidian distance. 

Once a test statistic has been chosen, its sampling distribution needs
to be derived. When the sampling distribution is not known in closed
form (either exactly or approximately) it can be calculated
empirically using, for example, a delete-1 jackknife \cite{shao}. Let
$d_i = D(H_i, H_{a,-i})$ be the distance from the $i$th histogram to
the aggregated histogram of the $n-1$ remaining realisations, denoted
by $H_{a,-i}$. Then, the empirical CDF of the test statistic is
\begin{equation*}
F(d) = p(D \le d) \approx \frac{1}{n}\sum_{i = 1}^n 1_{[[d_i<d]]},
\end{equation*}
where $1_{[[\cdot]]}=1$ if $[[\cdot]]$ is true and zero otherwise. Let
$d_{1-\alpha}$ be the rejection threshold corresponding to the
$(1-\alpha)$ quantile of the CDF. When $H_{data}$ is observed I can
calculate the distance from the data to the full aggregate histogram
$H_{a}$:
\begin{equation*}
d_{data} = D(H_{data}, H_{a})
\end{equation*} 
and then compare $d_{data}$ with $d_{1-\alpha}$.  If
$d_{data}>d_{1-\alpha}$ I reject the hypothesis that the models
generating $H_1,\dots,H_n$ and $H_{data}$ are the same.  It can be
shown that this test has a Type I error of approximately $1-\alpha$
when $n$ is large.
A similar hypothesis test could also be constructed using either a
delete-$d$ jackknife or a bootstrap method \cite{shao}.

\subsection{Hypothesis test on simulated data}
Suppose I have $n=50$ realisations of final colony size from the full
model for the following parameter sets: attractive ($A=-10$),
repulsive ($A=10$), no force ($A=0$), and no movement ($A=0$ and
$P_1=1$), all with doubling time $T=300$. The aggregate distributions
are shown in Figure \ref{fig:sims_spreading3}. I will refer to these
as the reference models or distributions. In Table
\ref{tb:distancedistrib} I show the maximum, 95\% quantile, 75\%
quantile, and mean of the distributions of Euclidian distances
($\{d_i\}_{i=1}^n$) for these cases. The value of the 95\% quantile
(or similarly the $(0.95) n^{th}$ order statistic) serves as a
rejection threshold for hypothesis tests comparing new data with the
reference models.

\begin{table}[h!]
\begin{center}
\begin{tabular}{| l | c | c | c | c |}
\hline
Model & Max & 95\% Quantile & 75\% Quantile & Mean \\
\hline
Attractive ($A=-10$) & 16.59 & {\bf 12.94} & 11.10 & 9.67 \\
Repulsive ($A=10$) & 74.49 & {\bf 48.03} & 33.92 & 27.15 \\
Random ($A=0$) & 35.63 & {\bf 20.75} & 15.11 & 12.95 \\
No Movement & 17.65 & {\bf 17.23} & 13.02 & 10.52 \\
\hline
\end{tabular} 
\caption[Summary statistics of the distributions of distances $d_i$
for reference models]{Maximum, 95\% Quantile, 75\% Quantile, and Mean
  of the distributions of distances $d_i$ for 4 parameter settings of
  the full model (reference models). Parameters are as in
  \ref{tb:sim_M3}, with doubling time $T=300$, except for the no
  movement case which has stopping probability $P_1=1$.
  \label{tb:distancedistrib}}
\end{center}
\end{table}

I conducted four additional simulation runs, one for each of the
attractive, repulsive, random, and no movement cases. I then
calculated the Euclidian distance from each of these individual runs
to the reference histograms ($d_{data}$). These appear in Table
\ref{tb:distancetests}. The rows correspond to reference histograms,
and columns to individual runs. The distance from the individual run
to the reference distribution generated from the same set of
parameters is indicated in bold face. Notice that the bold value is
the lowest distance in each column. These bold distances are all lower
than the corresponding 95\% quantile noted in Table
\ref{tb:distancedistrib}, so I cannot reject the hypothesis that the
individual run is from the corresponding reference model. Distances to
all other reference distributions are higher than the respective 95\%
quantile. In these cases, the hypothesis that the individual run is
from the same distribution as the reference model is rejected. 

\begin{table}[h!]
\begin{center}
\begin{tabular}{| l | c | c | c | c |}
\hline
Model & Attractive & Repulsive & Random & No Movement \\
\hline
Attractive ($A=-10$) & {\bf 6.81} & 894.0 	&  49.16 	& 63.61 \\
Repulsive ($A=10$) &102.5 	& {\bf 23.35} & 166.3 & 189.5 \\
Random ($A=0$) 	& 27.74 	& 877.8 	& {\bf 16.76} & 40.76 \\
No Movement 		& 43.52 	& 1020 	& 58.20 	&  {\bf 16.80} \\
\hline
\end{tabular} 
\caption[Distances from new individual simulations to the reference
distributions.]{Distances from new individual simulations to the
  aggregate distributions ($d_{data}$). The new data sets are
  simulated using the same parameter settings as the four reference
  models. \label{tb:distancetests}}
\end{center}
\end{table}

The first set of comparisons presented above indicates that the method
is well-calibrated because we accept and reject where we expect to.
However, there is still the possibility that we would not be able to
distinguish between models that have similar (but not exactly the
same) parameters. To explore this, I ran ten additional simulations,
this time with at least one of either the force constant ($A$) or the
the doubling time ($T$) different from any of the reference models.
First, I looked at variation in the force constant only. In Table
\ref{tb:distancetests2} I show distances between the reference models
and individual runs with five different values of $A$, ranging from
$A=5$ (mild repulsion) through $A=-8$ (moderate attraction). Notice
that all of these distances are in the rejection region. Moreover,
most of them are also significantly larger than maximum of the within
model distances, $d_i$.
\begin{table}[h!]
\begin{center}
\begin{tabular}{| l | c | c | c | c | c |}
\hline
Model 	             & $A=5$ & $A=0.5$ & $A=-2$ & $A=-5$ & $A=-8$ \\
\hline
Attractive ($A=-10$) & 582.5 & 40.17 & 65.57 & 22.47 & 58.80 \\
Repulsive ($A=10$)   & 131.0 & 183.4 & 156.9 & 135.1 & 129.3 \\
Random ($A=0$) 	     & 559.6 & 68.00 & 92.15 & 25.34 & 81.63 \\
No Movement 	     & 682.4 & 97.07 & 110.1 & 35.83 & 99.00 \\
\hline
\end{tabular} 
\caption[Distances from new individual simulations to the reference
distributions: varying $A$]{Distances from new individual simulations
  to the aggregate distributions ($d_{data}$). The new data sets are
  simulated using the same parameter settings as the four reference
  models, except for the force constant $A$.
  \label{tb:distancetests2}}
\end{center}
\end{table}

Next, I changed the doubling time, so that $T=250$ and calculated the
distances from these new runs to the reference distributions (Table
\ref{tb:distancetests3}). Here again, all of the distances fall
outside of the non-rejection region, except for the individual repulsive
($A=10$) run (Table \ref{tb:distancetests3} in bold face). This is
likely because the individual cells move for much longer and form
smaller groups regardless of the doubling time, so distinguishing
between two repulsive cases with similar doubling times is likely to
be difficult.

\begin{table}[h!]
\begin{center}
\begin{tabular}{| l | c | c | }
\hline
Model & $A=-10$,  &$A=10$,  \\
 & $T=250$ & $T=250$\\
\hline
Attractive ($A=-10$) & 42.65 & 1169  \\
Repulsive ($A=10$) & 102.0 & {\bf 38.76}\\
Random ($A=0$) & 61.39 & 1152  \\
No Movement & 78.20 & 1344  \\
\hline
\end{tabular} 
\caption[Distances from new individual simulations to the reference
distributions: varying $T$]{Distances from new individual simulations
  to the aggregate distributions ($d_{data}$). The new data sets are
  simulated using the same parameter settings as the attractive and
  repulsive reference models, except with $T=250$.
  \label{tb:distancetests3}}
\end{center}
\end{table}

Finally, I calculated distances between individual runs and the
reference distributions (with both $A$ and $T$ different from those
used to generate the aggregate distributions -- Table
\ref{tb:distancetests4}). Again, all of these distances are large
enough to reject the hypothesis that the individual runs were generated from
the model corresponding to the aggregate distributions.

\begin{table}[h!]
\begin{center}
\begin{tabular}{| l | c | c |  c |}
\hline
Model & $A=-5$, &$A=5$, & $A=60$, \\
& $T=250$ &$T=250$ & $T=450$ \\
\hline
Attractive ($A=-10$) & 64.21 & 649.3 & 578.4 \\
Repulsive ($A=10$) & 139.1 & 200.3 & 51.02\\
Random ($A=0$) & 88.89 & 622.7  & 574.6\\
No Movement & 107.33 & 775.6  & 663.8 \\
\hline
\end{tabular} 
\caption[Distances from new individual simulations to the reference
distributions: varying $A$\& $T$]{Distances from new individual
  simulations to the aggregate distributions ($d_{data}$). The new
  data sets are simulated using the same parameter settings as the
  four reference models, except for $A$ and $T$ (as indicated in the
  table). \label{tb:distancetests4}}
\end{center}
\end{table}

\section{Fitness of bacteria in microcolonies}\label{predation}
This analysis of models of microcolony formation presented thus far
neglects the question: why do bacteria employ a strategy that results
in a particular microcolony size? We would expect that the strategies
must be related to the fitness of the bacteria. For instance, if
bacteria repel each other, there must be an advantage to small group
size, at least initially, perhaps because of better access to
nutrients and increased reproductive rates.  On the other hand,
aggregation would provide some advantage that favours large groups,
such as protection from predation \cite{matz:2005}.

Mortality rates are often size dependent
\cite{lorenzen:1996,mcgurk:1986a}. For instance, a traditional model
for a size dependent mortality rate, $m(L)$, is
\begin{equation}
m(L) = m_1 + \frac{m_2}{L}\label{eq:group_m}
\end{equation}
where the length of the organism is $L$, $m_1$ is some constant
mortality, and $m_1 + m_2$ gives the mortality rate of the organism at
length $L=1$. For a bacterial microcolony, I use this form for the
average mortality of a single bacterium in a group, where $L$ is the
diameter of the group (assuming the group is approximately circular).
Measuring the group size as a multiple of the diameter of a cell,
$m_1+m_2$ is the mortality rate of a single bacterium and  $m_1$ is the
mortality rate of bacteria in a large group.  

Living in a microcolony or biofilm may improve survival, but since
resources are limited, living in a group is likely to decrease the
rate of reproduction, or increase the doubling time. For instance I
assume a simple relationship between group size and doubling time,
$T$, such that
\begin{equation}
T(L) = T_1 + T_2 L^{\gamma}. \label{eq:group_T}
\end{equation}
I examine the case of $\gamma = 1$. In this case, $T_1+T_2$ is the
doubling time of a single bacterium. 

Johnson and Mangel (2006) propose a simple model of bacterial fitness
when bacteria age, {\it i.e.} experience limited reproduction. In
this model the fitness of a bacterium, $\tilde{r}$ is
\begin{align}
\tilde{r} &\approx \frac{\ln{2}}{T} -m -\frac{\epsilon}{T} \label{eq:colony_r}
\end{align}
where $m$ is the mortality rate, $T$ is the doubling time, and
$\epsilon = \frac{2^{-a_{max}}}{\ln{2} + a_{max} 2^{-a_{max}}}$, where
$a_{max}$ is the maximum number of times a bacterium can double. Using
Eqns. (\ref{eq:group_m}) and (\ref{eq:group_T}) for $m$ and $T$ in
Equation (\ref{eq:colony_r}) gives the fitness of the bacterium as a
function of the group size $L$
\begin{equation}
\tilde{r}(L) \approx \frac{b}{T_1 + T_2 L} - m_1 - \frac{m_2}{L}
\end{equation}
where $b=\ln{2}-\epsilon$. The optimal group length $L^*$ will
maximise the fitness, so that $\frac{d\tilde{r}(L^*)}{dL} = 0$.
Solving for $L^*$ gives
\begin{equation*}
L^*_\pm = \frac{-T_1 \pm T_1\left(\frac{b}{m_2T_2}\right)^{\frac{1}{2}}}{T_2-\frac{b}{m_2}}.
\end{equation*}
For this to make sense as a length, $L^*$ must be positive. The first
of these solutions
\begin{equation*}
L^*_+ = \frac{-T_1 + T_1\left(\frac{b}{m_2T_2}\right)^{\frac{1}{2}}}{T_2-\frac{b}{m_2}}
\end{equation*}
cannot yield positive group lengths. The second solution 
\begin{align}
  L^*_- &= \frac{-T_1 -
    T_1\left(\frac{b}{m_2T_2}\right)^{\frac{1}{2}}}{T_2-\frac{b}{m_2}}\nonumber \\
  &= \frac{T_1 +
    T_1\left(\frac{b}{m_2T_2}\right)^{\frac{1}{2}}}{\frac{b}{m_2}-T_2} \label{eq:L_star1}
\end{align}
corresponds to a positive solution if $T_2<\frac{b}{m_2}$. If I define
$\omega = \frac{b}{m_2 T_2}$, Equation (\ref{eq:L_star1}) becomes
\begin{equation}
L^* = \frac{T_1}{T_2} \left( \frac{\omega^{\frac{1}{2}}+1}{\omega-1} \right)  \label{eq:L_star}
\end{equation}
where $\omega > 1$. Notice that the optimal group size does not depend
upon the size independent mortality rate $m_1$, since this only shifts
the fitness by a constant.

If the group of bacteria is approximately circular, and the area
covered by the groups is approximately the number of cells in the group,
$N$, multiplied by the area covered by a single cell, then the
diameter of the group scales as $N^{\frac{1}{2}}$. The optimal number
of individuals in a group is then approximately
\begin{equation}
N^* \approx (L^*)^2 = \left(\frac{T_1}{T_2} \left( \frac{\omega^{\frac{1}{2}}+1}{\omega-1} \right) \right)^2 .\label{eq:N_star}
\end{equation}

Very large values of $N^*$ will be optimal if at least one of two
conditions holds. The first is that $T_2 << T_1$, {\it i.e.}, larger
group sizes only have a small effect on the doubling rate, so that the
ratio of $T_1$ to $T_2$ will be large. The optimal group size will
also be large when $\omega \approx 1$ which corresponds to $m_2
\approx \frac{b}{T_2}$. This means that if the doubling time increases
quickly with group size, {\it i.e.} $T_2$ is large, the group must
provide significant protection for large group sizes to be optimal. If
$m_2$ is due to predation, for instance, then the threat would not be
constant in time, and the optimal group size would depend on predators
being present in the environment. This requires that bacteria be able
to sense that there is danger and modify their behaviour accordingly.
{\it V. cholerae} appear to exhibit this behaviour. Matz {\it {\it et
    al.}}  \cite{matz:2005} studied biofilm formation in {\it V.
  cholerae} in the presence of predatory protozoa.  They found that
bacteria living in biofilms are protected from protozoa compared to
those living in planktonic state, and that bacteria in the biofilm
produce compounds which inhibit the growth of the protozoa. They also
found that if there is predation of the planktonic state, biofilm
formation is significantly enhanced compared to growth without
predation. The bacteria appear to regulate these various responses to
the protozoa via quorum sensing.

The calculation of optimal group size here gives a way to estimate how
strong predation must be in order for the biofilm state to be
preferred, and estimate what levels of predation and what group sizes
are necessary to initiate quorum sensing. On the other hand, if the
mortality rates of individual bacteria and large groups can be
measured and the doubling time for planktonic bacteria is known (as in
the experiments by Matz {\it {\it et al.}}), the values of $T_2$
required to select for large group sizes could be calculated. These
calculations also indicate what types of environments
(presence/absence of predators, nutrient limitation, {\it etc.}) are
likely to be correlated with the different types of surface movements
explored in the IBM introduced in this paper. For instance, if $T_2$
is large, and the risk of predation is low ($m_2$ is small), then
repulsive forces, which results in smaller group sizes, would be
predicted to be prevalent. However, attraction between individuals,
which would allow larger groups to be formed faster, would be optimal
if the risk of predation is high.

\begin{figure}
\begin{center}\begin{tabular}{cc}
\includegraphics[trim = 30 0 0 30, scale=0.425]{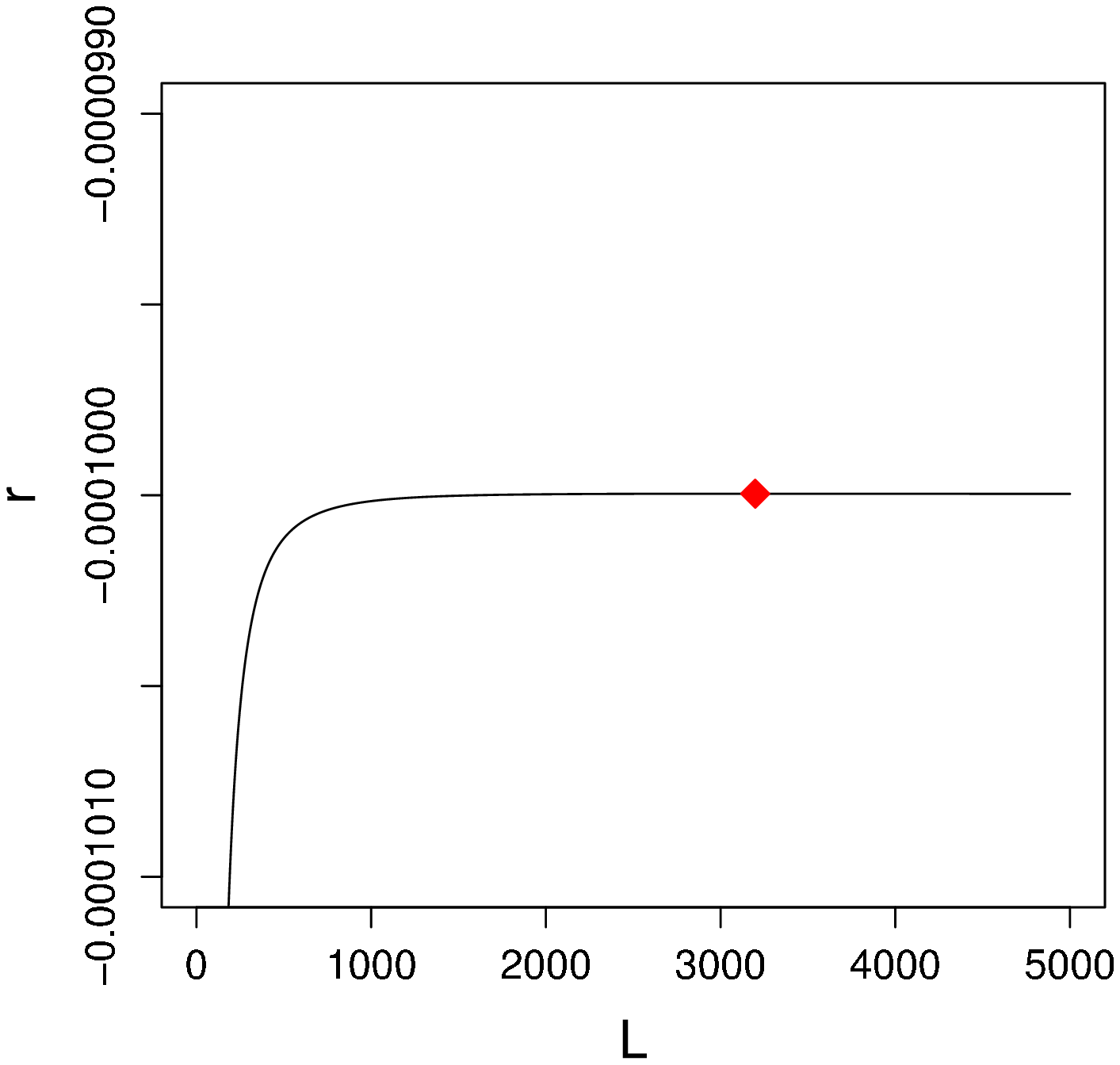} &
\includegraphics[trim = 10 0 0 30, scale=0.425]{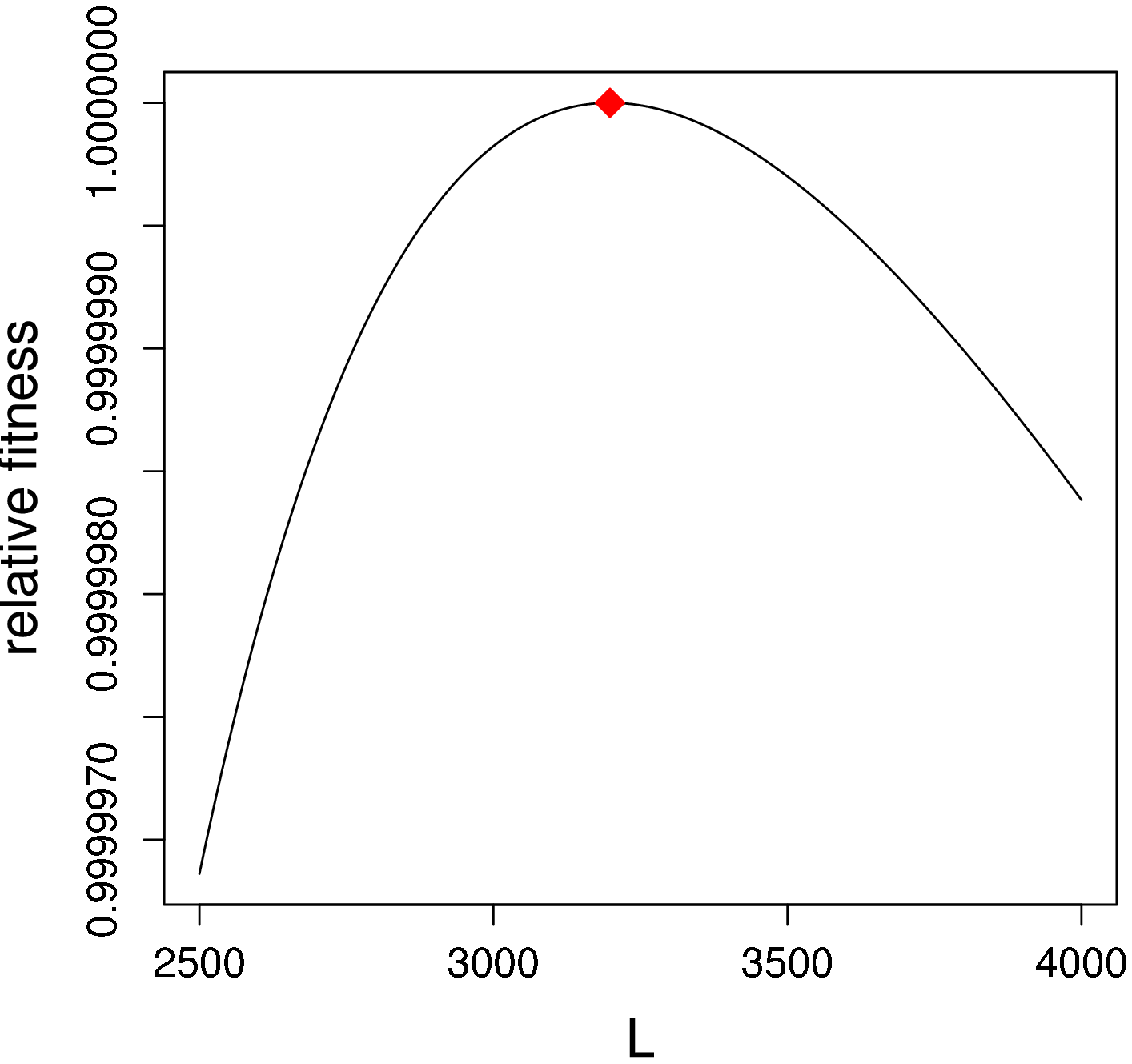} \\
{\small (a) } & {\small (b) }\\
\end{tabular}
\begin{tabular}{c}
\includegraphics[trim = 0 0 0 0, scale=0.425]{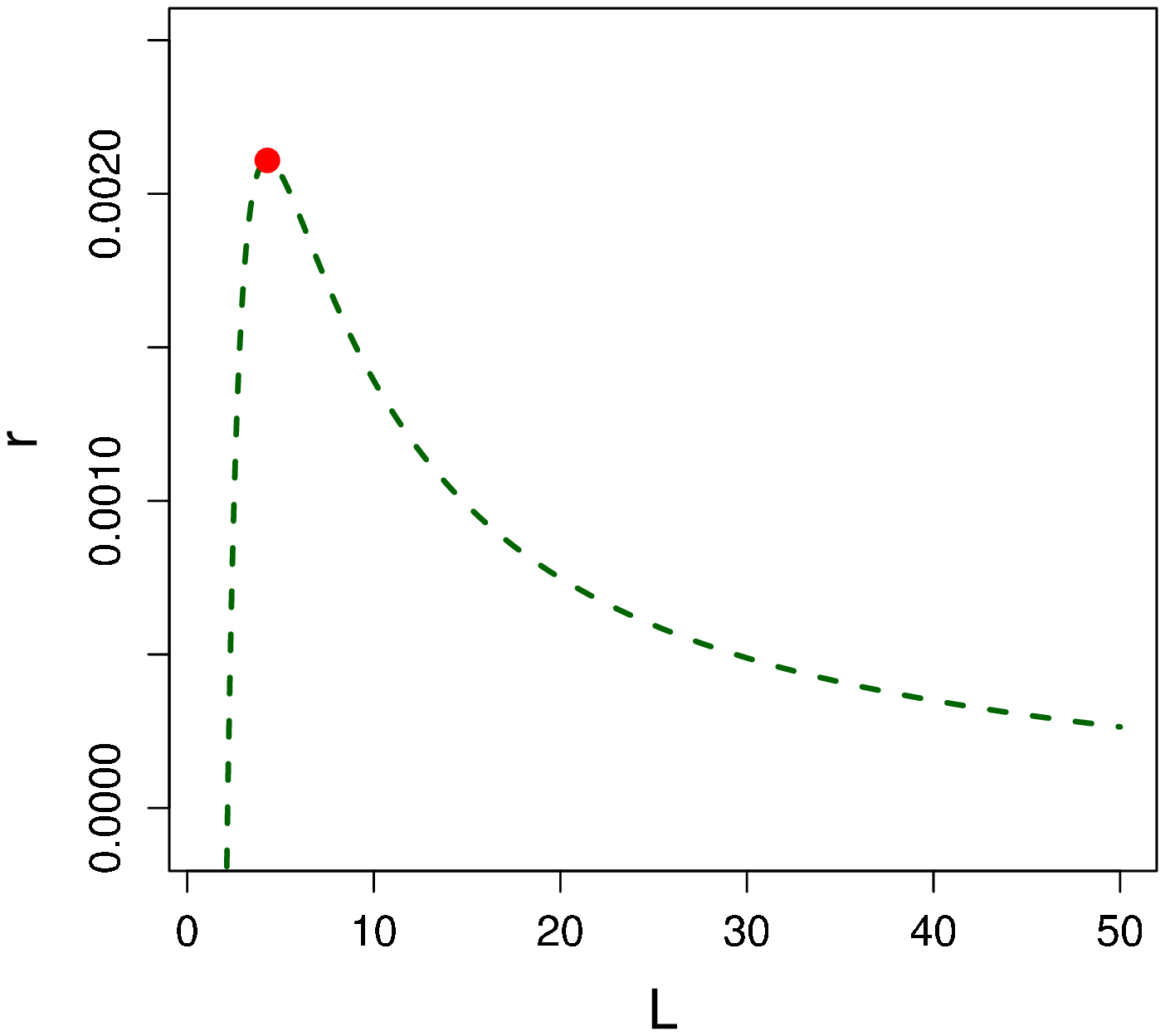} \\
{\small (c) } \\
\end{tabular}

\end{center}
\caption[Fitness of bacteria as a function of groups size (a).]{(a)
  Fitness of bacteria. $\tilde{r}$ as a function of groups size, $L$
  for parameters estimated from data in Matz, {\it et al.} $m_2 =
  2.079$ with $T_2 = 0.3334$, $m_1 = 0.0001$, $T_1 = 0.0068$,
  $b=\log2$. $L^*$ is indicated with a red point. (b) Close-up of (a)
  near the optimal group size, $L^*$.  The y-axis here indicates
  relative fitness Relative fitness ($\tilde{r}/\tilde{r}_{max}$) (c)
  $r$ vs. $L$ for parameters estimated from data in Matz, {\it et
    al.}, except with $m_2 = 2.06$.} \label{fig:r_vs_L_1}
\end{figure}

To make this more concrete, I calculate approximate values of the
parameters using the experimental data from \cite{matz:2005}. In
experiments lasting for 72 hours, Matz {\it et al.} found that with a
planktonic predator, the population of planktonic bacteria was reduced
by $>94\%$. On the other hand, the biofilm population was unaffected
by a surface predator. I assume a doubling time for planktonic
bacteria ($T_1 +T_2$) of $0.34$ hours (about 20 minutes). Since $T_2$
is unlikely to be very small (as biofilms tend to decrease the
diffusion of nutrients to many of the members) and the system favours
very large groups when predators are present (as was indicated by
increased biofilm formation in the presence of the planktonic
predator), I also assume that $T_2 \approx \frac{\ln{2}}{m_2} -
small$, so that $L^*$ is very large. The estimated values of the
parameters are then: $m_1\approx 0.0001$, $m_2 \approx 2.079$,
$T_1\approx 0.0068$, and $T_2 \approx 0.3332$, corresponding to an
optimal group size of $L^* = 3200$ or $N^* \approx 10^7$. In Figure
\ref{fig:r_vs_L_1} a \& b I show $\tilde{r}$ as a function of $L$ for
these parameters, with the optimal group size indicated on the curve.
Notice that for this case, the fitness is always less than one, and
the population will decline, whereas if $m_2$ is reduced a small
amount (Figure \ref{fig:r_vs_L_1} c) the fitness would be positive.
This is because of the value of the constant mortality $m_1$. If this
mortality were zero, then the population in a large group would be
stationary ($\tilde{r}=0$).

\begin{figure}[h!]
\begin{center}\begin{tabular}{c}
\includegraphics[trim = 10 0 0 0, scale=0.475]{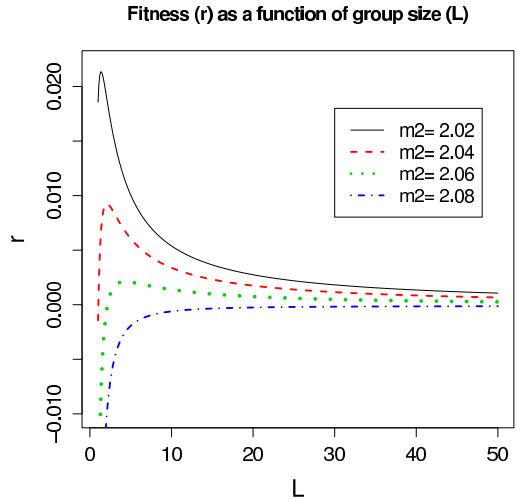} \\
{\small (a)}\\
\includegraphics[trim = 10 0 0 0, scale=0.475]{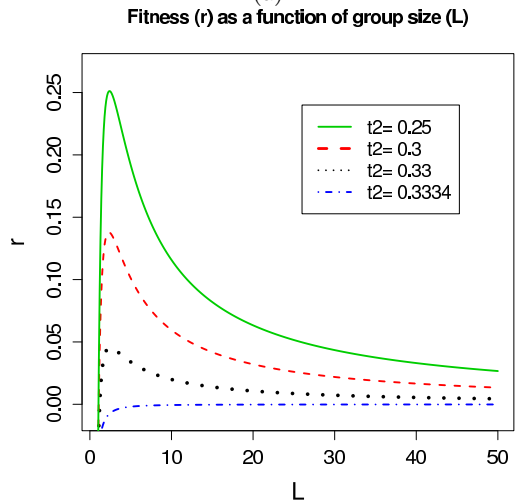} \\
{\small (b) }\\
\end{tabular}
\end{center}
\caption[Fitness of bacteria as a function of groups size. ]{Fitness
  of bacteria as a function of groups size for (a) 4 values of $m_2$
  with $T_2 = 0.3334$ and (b) 4 values of $T_2$ with $m_2 = 2.08$.
  Values of other parameters are: $m_1 = 0.0001$, $T_1 = 0.0068$,
  $b=\log2$.} \label{fig:r_vs_L_2}
\end{figure}

In Figure \ref{fig:r_vs_L_2} I show plots of $\tilde{r}$ as a function
of $L$ varying $m_2$ (Figure \ref{fig:r_vs_L_2} a) and varying $T_2$
(Figure \ref{fig:r_vs_L_2} b) with the other parameter values set to
those estimated here. For the optimal group size to be reduced to $N^*
\approx 1$, the predation mortality must be decreased to about
$m_2=2.02$, a reduction of only $\approx$ 3\%. In otherwords, this
model predicts that a small reduction in mortality of planktonic
bacteria would result in individual bacteria being more prevalent (top
three lines in Figure \ref{fig:r_vs_L_2} a). If the mortality, $m_2$,
remains constant, a decrease in the size dependent mortality rate to
$T_2 \approx 0.25$ gives a smaller optimal groups size (in this case
$N^* \approx 4$), so small groups should be more prevalent. Notice
that when the optimal group size is small the fitness surface is
sharply peaked around the optimal value, so there would be little
variability in clump sizes. On the other hand, as the optimal group
size increases, the fitness surface flattens out. This is especially
obvious in Figure \ref{fig:r_vs_L_1} b, where the difference in
fitness between the maximum fitness at $L^*=3200$ and the fitness of a
groups of size $L=2500$ is on the order of $10^{-9}$. This indicates
that there would be more variability in clump sizes under these
conditions, since even groups with sizes that are far from the optimal
will have similar fitness.

\section{Discussion}
Individual Based Models have traditionally been used to explore how
the behaviour of individuals combine to create large scale or group
patterns. Instead of encoding a particular large scale pattern
explicitly into the dynamics of the system, IBMs seek to identify
individual behaviours that can generate emergent properties or
patterns.

The formation of bacterial communities, such as biofilms, has major
implications for survival and fitness of bacteria. To understand
patterns of group formation in these organisms we must gather
information about how behaviour of individual bacteria influences these
patterns. The IBM presented here is one way to approach this
problem. 

This model provides useful qualitative and quantitative predictions of
how the size of microcolonies depends on the behaviour of the cells.
Qualitative results include insight into how inter-cellular
interactions tend to influence the mean colony size, maximum colony
size, and heaviness of distribution tails. For example, repulsive
forces result in smaller mean colony sizes, with short tails, whereas
attractive forces increase the likelihood of larger colonies, as well
the maximum observed colony size. We also expect that if predators are
present, individuals should aggregate more quickly. 

Quantitatively, the model predicts that the spacing between peaks is
proportional to the expected number of doublings in the observation
period. The results of this model also indicate that although the
patterns of colony size are complicated, different behaviours result in
patterns that are statistically different. This enables us to conclude
that groups of individuals are exhibiting different behaviour simply by
comparing the patterns of group sizes. The distinctness of the
patterns means that this model would be useful for making quantitative
predictions, and for inferring from empirical data the strategies
employed by bacteria in the initial stages of biofilm formation.

When studying bacterial biofilm formation, we are often interested in
genes that are important for the development of the biofilms. In
laboratory experiments, genes that are thought to be involved in
attachment or biofilm formation are deleted or mutated. Bacteria with
these modifications are then grown {\it in vitro} to see if they can
still form biofilms or produce certain substances. However, often the
particular functions of the genes involved are unknown. Even if the
function is known, why that gene is important or how it influences
community structure may be unknown. For instance, genes that are
believed to influence chemotaxis and motility in {\it V. cholerae} are
also important to biofilm formation. However, how these genes mediate
inter-cellular interactions and biofilm formation is well not
understood. As a first step, we are interested in learning if these
genes may be playing a part in the initial stages of biofilm
formation. The model presented here provides a framework for exploring
these kinds of questions.

For instance, imagine a flow cell experiment. The flow cell is first
inoculated with bacteria, which are allowed a short amount of time to
form initial attachments to the surface. After a short time has
elapsed, the flow cell is flushed so that any bacteria not already
attached to the surface are removed. Initial images could be taken to
determine the approximate initial cell density, although this is
likely not necessary, unless the initial densities are either very
high or very low. Then fresh medium is allowed to flow through the
flow cell. The remaining bacteria reproduce and/or move around on the
surface. Images can be taken at later times, and the approximate
number of cells per colony can be recorded. These data can then be
compared with model predictions. This would provide information about
which of the models presented best matches the results of the
experiment as well as what type of behaviour the bacteria are
exhibiting, {\it i.e.} if they are attracting via chemotaxis, or
moving along the surface at all.  Experiments involving bacteria with
certain mutations, such as one that completely removes the ability of
a bacteria to produce a flagellum or a mutation that alters a
chemotaxis system, could be used to pin down estimates of parameters
such as the doubling time or effective force of attraction/repulsion,
and indicate which genes are correlated with specific behaviours.

Before comparing the simulated distributions with empirical data, it
could be useful to estimate empirically the probability of making a
type II error ({\it i.e.} accepting the hypothesis that the data is
generated by a particular model, when the data was actually generated
by another model). This would require simulating large sets of data
from each model, and then comparing the ``data'' with the various
reference distributions using the jackknife type hypothesis test
described earlier.

Fairly simple extensions and modifications of the models explored here
could be used to explore other kinds of bacterial communities. For
instance, by adjusting the shoving parameters, the clumpiness and
density of the communities can be manipulated, which may give insight
into what behaviours are involved in determining colony morphology.
Allowing shoving to shape the surface attached groups in three
dimensions could also be useful in this regard. One way to do this
would be to combine the IBM explored here with BacSim (or some other
biofilm modelling tool) to explore how colony and biofilm structures
are influenced by the behaviours of individuals during initial surface
colonisation.

Another important extension would be to explore how different
functional forms for the interaction forces between individuals
influences the distributions of colony sizes. In particular, modelling
chemotaxis explicitly together with an analysis of what functional
form of direct forces is the best approximation for chemotaxis would
be useful.

The model could also be easily expanded to include multiple types of
bacteria, or genotypic and phenotypic variation between individuals.
For instance, one study on {\it Pseudomonas aeruginosa}, conducted by
Klausen, {\it et al.} (2003) looked at aggregation and biofilm
formation when individuals were marked with one of two florescent
markers (either blue or green). They concluded that because bacterial
colonies were not 50-50 blue-green, the bacteria were not aggregating
on the surface. The IBM presented here could be used to explore how
the size and composition of groups varies depending on how individuals
interact with other individuals (of the same and different colors)
more explicitly, in order to draw more concrete conclusions about the
behaviour of these bacteria.

\section{Acknowledgements}
This work was supported by a UC Presidents Dissertation Year
Fellowship. Thanks to Marc Mangel for comments on earlier drafts and to
Robert Gramacy for advice on statistical methods.

\nocite{johnson:2006}
\singlespace
\bibliographystyle{apalike}
\bibliography{cholera,biofilm,disease,other,biology,math}

\begin{thebibliography}{}

\bibitem[Branda et~al., 2005]{branda:2005}
Branda, S.~S., Vika, {\AA}., Friedmanb, L., and Kolter, R. (2005).
\newblock {Biofilms: the matrix revisited}.
\newblock {\em TRENDS in Microbiology}, 13(1):20--26.

\bibitem[Davey and O'Toole, 2000]{davey:2000}
Davey, M.~E. and O'Toole, G.~A. (2000).
\newblock {Microbial Biofilms: from Ecology to Molecular Genetics}.
\newblock {\em Microbiol. Mol. Biol. Rev.}, 64(4):847--867.

\bibitem[Davies et~al., 1998]{davies:1998}
Davies, D.~G., Parsek, M.~R., Pearson, J.~P., Iglewski, B.~H., Costerton,
  J.~W., and Greenberg, E.~P. (1998).
\newblock {The Involvement of Cell-to-Cell Signals in the Development of a
  Bacterial Biofilm}.
\newblock {\em Science}, 280(5361):295--298.

\bibitem[Grimm and Railsback, 2005]{grimm}
Grimm, V. and Railsback, S.~F. (2005).
\newblock {\em {Individual-based Modeling and Ecology}}.
\newblock Princeton University Press, Princeton, NJ.

\bibitem[Gueron and Levin, 1995]{gueron:1995}
Gueron, S. and Levin, S.~A. (1995).
\newblock {The Dynamics of Group Formation}.
\newblock {\em Mathematical Biosciences}, 128:243--264.

\bibitem[Gueron et~al., 1996]{gueron:1996}
Gueron, S., Levin, S.~A., and Rubenstein, D.~I. (1996).
\newblock {The Dynamics of Herds: From Individuals to Aggregations}.
\newblock {\em Journal of Theoretical Biology}, 182:85--98.

\bibitem[Hahn and H{\"{o}}fle, 2001]{hahn:2001}
Hahn, M.~W. and H{\"{o}}fle, M.~G. (2001).
\newblock Grazing of protozoa and its effect on populations of aquatic
  bacteria.
\newblock {\em FEMS Microbiology Ecology}, 35:113--121.

\bibitem[Heydorn et~al., 2002]{heydorn:2002}
Heydorn, A., Ersboll, B., Kato, J., Hentzer, M., Parsek, M.~R., Tolker-Nielsen,
  T., Givskov, M., and Molin, S. (2002).
\newblock {Statistical Analysis of Pseudomonas aeruginosa Biofilm Development:
  Impact of Mutations in Genes Involved in Twitching Motility, Cell-to-Cell
  Signaling, and Stationary-Phase Sigma Factor Expression}.
\newblock {\em Appl. Environ. Microbiol.}, 68(4):2008--2017.

\bibitem[Jefferson, 2004]{jefferson:2004}
Jefferson, K.~K. (2004).
\newblock {What drives bacteria to produce a biofilm?}
\newblock {\em FEMS Microbiology Letters}, 236:163--173.

\bibitem[Johnson and Mangel, 2006]{johnson:2006}
Johnson, L.~R. and Mangel, M. (2006).
\newblock Life histories and the evolution of aging in bacteria and other
  single-celled organisms.
\newblock {\em Mechanisms of Ageing and Developement}, 127(10):786--793.

\bibitem[Keller and Segel, 1970]{keller:1970}
Keller, E.~F. and Segel, L.~A. (1970).
\newblock {Initiation of Slime Mold Aggregation Viewed as an Instability}.
\newblock {\em Journal of Theoretical Biology}, 26:399--415.

\bibitem[Keller and Segel, 1971a]{keller:1971a}
Keller, E.~F. and Segel, L.~A. (1971a).
\newblock {Model for Chemotaxis}.
\newblock {\em Journal of Theoretical Biology}, 30:225--234.

\bibitem[Keller and Segel, 1971b]{keller:1971b}
Keller, E.~F. and Segel, L.~A. (1971b).
\newblock {Traveling Bands of Chemotactic Bacteria: A Theoretical Analysis}.
\newblock {\em Journal of Theoretical Biology}, 30:235--248.

\bibitem[Klausen et~al., 2003]{klausen:2003}
Klausen, M., Heydorn, A., Ragas, P., Lambertsen, L., Aaes-J{\o}rgensen, A.,
  Molin, S., and Tolker-Nielsen, T. (2003).
\newblock {Biofilm formation by {\it Pseudomonas aeruginosa} wild type,
  flagella and type IV pili mutants}.
\newblock {\em Molecular Microbiology}, 48(6):1511--1524.

\bibitem[Kramers, 1940]{kramers:1940}
Kramers, H.~A. (1940).
\newblock {Brownian Motion in a Field of Force and the Diffusion Model of
  Chemical Reations}.
\newblock {\em Physica}, 7(4):284--288.

\bibitem[Kreft et~al., 1998]{kreft:1998}
Kreft, J.-U., Booth, G., and Wimpenny, J. W.~T. (1998).
\newblock {BacSim, a simulator for individual-based modeling of bacterial
  colony growth}.
\newblock {\em Microbiology}, 144:3275--3287.

\bibitem[Kreft et~al., 2001]{kreft:2001}
Kreft, J.-U., Picioreanu, C., Wimpenny, J. W.~T., and van Loosdrecht, M. C.~M.
  (2001).
\newblock {Individual-based modeling of biofilms}.
\newblock {\em Microbiology}, 147:2897--2912.

\bibitem[Lee et~al., 2001]{lee:2001}
Lee, C., Hoopes, M., Diehl, J., Gilliland, W., Huxel, G., Leaver, E., McCann,
  K., Umbanhowar, J., and Mogilner, A. (2001).
\newblock {Non-local Concepts and Models in Biology}.
\newblock {\em Journal of Theoretical Biology}, 210:201--219.

\bibitem[Lorenzen, 1996]{lorenzen:1996}
Lorenzen, K. (1996).
\newblock The relationship between body weight and natural mortality in
  juvenile and adult fish: a comparison of natural ecosystems and aquaculture.
\newblock {\em Journal of Fish Biology}, 49:627--647.

\bibitem[Madigan and Martinko, 2005]{brock}
Madigan, M. and Martinko, J. (2005).
\newblock {\em Brock Biology of Microorganisms (11th Edition)}.
\newblock Prentice Hall, Upper Saddle River, NJ.

\bibitem[Matz et~al., 2005]{matz:2005}
Matz, C., McDougald, D., Moreno, A.~M., Yung, P.~Y., Yildiz, F.~H., and
  Kjelleberg, S. (2005).
\newblock Biofilm formation and phenotypic variation enhance predation-driven
  persistence of {\it vibrio cholerae}.
\newblock {\em PNAS}, 102(46):498--506.

\bibitem[Mayer et~al., 1999]{mayer:1999}
Mayer, C., Moritz, R., Kirschner, C., Borchard, W., Maibaum, R., Wingender, J.,
  and Flemming, H.-C. (1999).
\newblock The role of intermolecular interactions: studies on model systems for
  bacterial biofilms.
\newblock {\em International Journal of Biological Macromolecules}, 26:3--16.

\bibitem[McGurk, 1986]{mcgurk:1986a}
McGurk, M.~D. (1986).
\newblock Natural mortality of marine pelagic fish eggs and larvae: role of
  spacial patchiness.
\newblock {\em Marine Ecology}, 43:227--242.

\bibitem[Mogilner et~al., 2003]{mogilner:2003}
Mogilner, A., Edelstein-Keshet, L., Bent, L., and Spiros, A. (2003).
\newblock Mutual interactions, potentials, and individual distance in a social
  aggregation.
\newblock {\em Journal of Mathematical Biology}, 47(4):353--389.

\bibitem[Okubo, 1986]{okubo:1986}
Okubo, A. (1986).
\newblock Dynamical aspects of animal grouping: Swarms, schools, flocks, and
  herds.
\newblock {\em Advances in Biophysics}, 22:1--94.

\bibitem[Parsek and Greenberg, 2005]{parsek:2005}
Parsek, M.~R. and Greenberg, E. (2005).
\newblock {Sociomicrobiology: the connections between quorum sensing and
  biofilms}.
\newblock {\em TRENDS in Microbiology}, 13(1):27--33.

\bibitem[Picioreanu and van Loosdrecht, 2003]{picioreanu}
Picioreanu, C. and van Loosdrecht, M. C.~M. (2003).
\newblock {\em Use of mathematical modelling to study biofilm development and
  morphology}, pages 5--15.
\newblock IWA, London.

\bibitem[Picioreanu et~al., 1999]{picioreanu2}
Picioreanu, C., van Loosdrecht, M. C.~M., and Heijnen, J. (1999).
\newblock {\em Multidimensional modelling of biofilm structure}.
\newblock Atlantic Canada Society for Microbial Ecology, Halifax, Canada.

\bibitem[Picioreanu et~al., 1998]{picioreanu:1997}
Picioreanu, C., van Loosdrecht, M. C.~M., and Heijnen, J.~J. (1998).
\newblock {Mathematical Modeling of Biofilm Structure with a Hybrid
  Differential-Discrete Cellular Automaton Approach}.
\newblock {\em Biotechnology and Bioengineering}, 58(1):101--116.

\bibitem[Roszak and Colwell, 1987]{roszak:1987}
Roszak, D.~B. and Colwell, R.~R. (1987).
\newblock Survival strategies of bacteria in the natural environment.
\newblock {\em Microbiological Reviews}, 51(3):365--379.

\bibitem[Shao and Tu, 1995]{shao}
Shao, J. and Tu, D. (1995).
\newblock {\em {The Jackknife and Bootstrap}}.
\newblock {Springer Series in Statistics}. Spinger-Verlag, New York, NY.

\bibitem[Suntharalingam and Cvitkovitch, 2005]{suntharalingam:2005}
Suntharalingam, P. and Cvitkovitch, D.~G. (2005).
\newblock {Quorum sensing in streptococcal biofilm formation}.
\newblock {\em TRENDS in Microbiology}, 13(1):3--6.

\bibitem[Tyutyunov et~al., 2004]{tyutyunov:2004}
Tyutyunov, Y., Senina, I., and Arditi, R. (2004).
\newblock {Clustering due to Acceleration in the Response to Population
  Gradient: A Simple Self-Organization Model}.
\newblock {\em The American Naturalist}, 164(6):722--735.

\bibitem[van Loosdrecht et~al., 2002]{vanLoosdrecht:2002}
van Loosdrecht, M. C.~M., Heijnen, J.~J., Eberl, H., Kreft, J.-U., and
  Picioreanu, C. (2002).
\newblock {Mathematical modelling of biofilm structures}.
\newblock {\em Antonie van Leeuwenhoek}, 81:245--256.

\bibitem[Watnick and Kolter, 1999]{watnick:1999}
Watnick, P.~I. and Kolter, R. (1999).
\newblock {Steps in the development of a {\it Vibrio cholerae} El Tor biofilm}.
\newblock {\em Mol Microbiol}, 34(3):586--586.

\bibitem[Wolfram, 2002]{wolfram}
Wolfram, S. (2002).
\newblock {\em {A New kind of Science}}.
\newblock Wolfram Media, Inc., Champaign, IL.

\bibitem[Yildiz and Schoolnik, 1999]{yildiz:1999}
Yildiz, F.~H. and Schoolnik, G.~K. (1999).
\newblock {Vibrio cholerae O1 El Tor: Identification of a Gene Cluster Required
  for the Rugose Colony Type, Exopolysaccharide Production, Chlorine
  Resistance, and Biofilm Formation}.
\newblock {\em Proc. Natl. Acad. Sci. USA}, 96(7):4028--4033.

\end{thebibliography}

\end{document}